\documentclass[a4paper,11pt]{article}
\pdfoutput=1 

\usepackage{amsfonts}
\usepackage{amsmath,bm}
\usepackage{amssymb,braket,dsfont,mathtools}
\usepackage{array}
\usepackage{bbold}
\usepackage[backend=biber,style=numeric-comp,subentry,sorting=none]{biblatex}
\usepackage{booktabs}
\usepackage[makeroom]{cancel}
\usepackage[T1]{fontenc}
\usepackage[margin=1in]{geometry}
\usepackage{graphicx,rotate,multicol,caption}
\usepackage[final]{hyperref}
\hypersetup{
	colorlinks=true,
	linkcolor=blue,
	citecolor=blue,
	filecolor=magenta,
	urlcolor=blue
}
\usepackage[utf8]{inputenc}
\usepackage{mathrsfs}
\usepackage{multirow,tabularx}
\usepackage{parskip}
\usepackage{slashed}
\usepackage{soul}
\usepackage{subcaption}
\usepackage{xcolor}
\newcommand{\beq}{\begin{equation}}
\newcommand{\eeq}{\end{equation}}
\newcommand{\bea}{\begin{eqnarray}}
\newcommand{\eea}{\end{eqnarray}}

\newcommand{\nn}{\nonumber}
\newcommand{\Fig}[1]{Fig.~\ref{#1}}

\newcommand{\Eq}[1]{Eq.~\eqref{#1}}
\newcommand{\Eqs}[2]{Eqs.~\eqref{#1}-\eqref{#2}}
\newcommand{\Tab}[1]{Table \ref{#1}}
\newcommand{\eV}{{\rm eV}}
\newcommand{\GeV}{{\rm GeV}}
\newcommand{\TeV}{{\rm TeV}}

\newcommand{\vep}{\varepsilon}

\newcommand{\vu}{\upsilon_u}
\newcommand{\e}{{\rm e}}
\newcommand{\f}{{\rm f}}
\newcommand{\BR}{\rm BR}

\newcommand{\MsL}{M_{\tilde \ell_L}^2}
\newcommand{\RHn}{RH$\nu$}
\newcommand{\ypmns}{Y_\nu^{\rm pmns}}
\newcommand{\yckm}{Y_\nu^{\rm ckm}}
\newcommand{\vpmns}{V_{\rm pmns}}
\newcommand{\vckm}{V_{\rm ckm}}
\newcommand{\gmu}{a_\mu}
\newcommand{\gmususy}{a_\mu^{\rm susy}}
\newcommand{\kGJ}{\rm k_{GJ}}

\newcommand{\mgut}{m_{\rm GUT}}

\defbibentryset{Ref_seesaw}{Minkowski:1977sc, Yanagida:1979as, Mohapatra:1979ia, GellMann:1980vs, 10.1007/978-1-4684-7197-7_15, Schechter:1980gr}
\defbibentryset{Ref_HP}{Wilson:1970ag, tHooft:1980xss}
\defbibentryset{Ref_SK}{Fukuda:1998mi, Fukuda:1998ub}
\defbibentryset{Ref_g2SM}{Blum:2018mom, Keshavarzi:2018mgv}
\defbibentryset{Ref_g2models}{Lindner:2016bgg, Cox:2018qyi, Cox:2018vsv, Abdughani:2019wai}
\defbibentryset{Ref_susydarkmatter}{Goldberg:1983nd, Ellis:1983ew}
\defbibentryset{Ref_susyGUT}{Dimopoulos:1981yj, Ellis:1990wk, Amaldi:1991cn, Langacker:1991an}
\defbibentryset{Ref_NUHMnat}{Chan:1997bi, Feng:1999mn, Feng:1999zg}
\defbibentryset{Ref_RNS125Higgs}{Baer:2012up, Baer:2012cf}
\defbibentryset{Ref_NUHM3}{Nath:1997qm, Lleyda:1993xf, Kobayashi:1994np}
\defbibentryset{Ref_HiggsinoWorld}{Kane:1996dd, KANE1998144}
\defbibentryset{Ref_HiggsinoPheno}{Han:2013usa, Abdughani:2017dqs, Ren:2017ymm, Wu:2017kgr, Han:2018rkz, Abdughani:2019wuv}
\defbibentryset{Ref_SUSY_RHn}{Borzumati:1986qx, Ilakovac:1994kj, Casas:2001sr}
\defbibentryset{Ref_NeutOthers}{deSalas:2017kay, Capozzi:2017ipn}
\defbibentryset{Ref_GJflavour}{Ross:2002fb, King:2003rf}
\defbibentryset{Ref_ourGJflavour}{Ross:2004qn, Das:2016czs, deMedeirosVarzielas:2005ax, Lopez-Ibanez:2017xxw, deMedeirosVarzielas:2017sdv, deMedeirosVarzielas:2018vab}
\defbibentryset{Ref_SusyHiggsUppBound}{Okada:1990vk, Okada:1990gg, Ellis:1990nz, Ellis:1991zd, Haber:1990aw}
\defbibentryset{Ref_LHCHIggs}{ATLAS:2012ae, Chatrchyan:2012tx}
\defbibentryset{Ref_ourHiggs}{Barenboim:2013bla, Barenboim:2013qya}
\begin{document}
\hfill {\small IFIC/20-10,\quad FTUV-20-0313}\\[0.5cm]

\begin{center}
	{\Large \bf \boldmath LFV and (g-2) in non-universal SUSY models with light higgsinos
	} \\[1cm]
	{\large
	  C. Han$^1$\footnote{hancheng@itp.ac.cn}
	, M.L. López-Ibáñez$^\dagger$\footnote{maloi2@uv.es}
	, A. Melis$^\ddagger$\footnote{aurora.melis@uv.es}
	, O. Vives$^\ddagger$\footnote{oscar.vives@uv.es}
	, L. Wu$^\star$\footnote{leiwu@itp.ac.cn}
	, J.M. Yang$^{\dagger,\, \S}$\footnote{jmyang@itp.ac.cn}}\\[20pt]
 	{\small
 	$^*$ School of Physics, KIAS, 85 Hoegiro, Seoul 02455, Republic of Korea. \\[5pt]
 	$^\dagger$ CAS Key Laboratory of Theoretical Physics, Institute of Theoretical Physics 
 	    \\ Chinese Academy of Sciences, Beijing 100190, P. R. China. \\[5pt]
 	$^\ddagger$ Departament de Física Tèorica, Universitat de València \& IFIC, Universitat
 	    de València \& CSIC, \\ Dr. Moliner 50, E-46100 Burjassot (València), Spain.\\[5pt]
 	$^\star$ Department of Physics and Institute of Theoretical Physics, Nanjing Normal    
 	    University \\ Nanjing 210023, P. R. China. \\[5pt]
 	$^\S$ School of Physical Sciences, University of Chinese Academy of Sciences,\\
 	    Beijing 100049, P. R. China
    }
\end{center}
\vspace*{1cm}
\begin{abstract}
\noindent 
 We consider a supersymmetric type-I seesaw framework with non-universal scalar masses at the GUT scale to explain the long-standing discrepancy of the anomalous magnetic moment of the muon. We find that it is difficult to accommodate the muon g-2 while keeping charged-lepton flavor violating processes under control for the conventional SO(10)-based relation between the up sector and neutrino sector. However, such tension can be relaxed by adding a Georgi-Jarlskog factor for the Yukawa matrices, which requires a non-trivial GUT-based model. In this model, we find that both observables are compatible for small mixings, CKM-like, in the neutrino Dirac Yukawa matrix.
\end{abstract}
\section{Introduction}
\label{sec:intro}
After the discovery of neutrino oscillations in 1998 by \texttt{Superkamiokande} \cite{Ref_SK}, the Standard Model (SM) was forced to include massive neutrinos. Yet, the smallness of their masses seems to require a new framework different from the SM Yukawa couplings in the charged lepton or quark sectors. Simultaneously, it was confirmed experimentally that neutrino mass eigenstates are a non-trivial combination of the flavour states.
Consequently, we know that the family lepton numbers, $L_\e,\, L_\mu$ and $L_\tau$, are violated in Nature.
This necessarily implies some degree of violation in the charged-lepton sector, although it has not yet been observed. 

Several mechanisms have been concocted to explain the extreme smallness of neutrino masses and all of them require the existence of new physics (NP). Unfortunately, the available experimental information on the mass splittings and mixing is still insufficient to disclose the physics behind their true origin. New observables are required to improve our understanding and charged-lepton-flavour violation (CLFV) is one of the best options at our reach. Nevertheless, the absence of any signal of CLFV may indicate that the associated NP is considerably heavy.
On the other hand, the persistent discrepancy between the experimental measurement of the anomalous magnetic moment of the muon \cite{Bennett:2006fi} and its SM prediction \cite{Ref_g2SM},
\beq
    \Delta a_\mu ~=~ a_\mu^{\rm exp} \:-\: a_\mu^{\rm SM} ~=~ \left(2.7\pm0.7\right) \times 10^{-9}, 
\eeq
remains an interesting motivation to explore new models \cite{Ref_g2models}.

With respect to neutrino masses, the type-I seesaw mechanism \cite{Ref_seesaw} seems to be the most natural extension of the SM to generate them.
It assumes the existence of right-handed neutrinos (RH$\nu$), which are singlets of the SM and, therefore, are allowed to have both Dirac and Majorana mass terms by the gauge symmetries.
Mediated by them, an effective dimension-5 Weinberg operator \cite{Weinberg:1979sa} would be induced, producing the light neutrino masses after the electroweak (EW) symmetry breaking.
Although RH$\nu$ may live at any scale between the EW and the GUT scale, a naïve dimensional analysis of the Weinberg operator signals a Majorana mass around the usual GUT scale $\sim 10^{15-16}\, \GeV$.
That means that the SM supplemented with RH$\nu$ at high energies %
suffers from a serious hierarchy problem \cite{Ref_HP}.

Despite the fact that no signal of superparticles has been found at LHC, supersymmetric models may still be the appropriate candidate to alleviate this problem and may also answer other questions, such as the nature of dark matter \cite{Ref_susydarkmatter} and the exact unification of the gauge couplings at the GUT scale \cite{Ref_susyGUT}.
In addition, the presence of right-handed neutrinos induces slepton mixing through the renormalization group equation (RGE), which may produce visible CLFV effects \cite{Ref_SUSY_RHn, Hisano:1995cp}.
In particular, SUSY models with non-universal scalar masses at the GUT scale %
seem to be favoured from naturalness considerations \cite{Ref_NUHMnat}, allowing for a Higgs boson mass at $125$ GeV and low electroweak fine-tuning \cite{Ref_RNS125Higgs}.
Here we will analyse supersymmetric non-universal Higgs models with an additional parameter for the third generation of scalar superpartners (NUHM3) \cite{Ref_NUHM3}.
We focus on the so-called light higgsino-world scenario \cite{Ref_HiggsinoWorld, Baer:2011ec, Ref_HiggsinoPheno, Han:2019vxi} in which the SUSY matter scalars are pushed into the multi-TeV scale while $\mu\lesssim 1\, \TeV$, as natural SUSY requires.

The paper is organised as follows: in Section \ref{sec:seesaw}, the main ingredients of a supersymmetric type-I seesaw model are presented.
We also comment on some specific features common to GUT-motivated scenarios.
In Section \ref{sec:RGE}, the effect of the right-handed neutrinos on the running of the slepton soft masses is described.
In Section \ref{sec:obs}, we discuss the main observables of our analysis providing some useful analytic relations between CLFV processes and $(g-2)_\mu$.
Section \ref{sec:analysis} is devoted to the results of our numerical scan.
We summarize our conclusions in Section \ref{sec:conclusion}.

\section{Supersymmetric type-I seesaw}
\label{sec:seesaw}
\begin{figure}[t!]
  \centering
    \begin{minipage}[c]{\textwidth}
        \centering
        \includegraphics[width=0.5\textwidth]{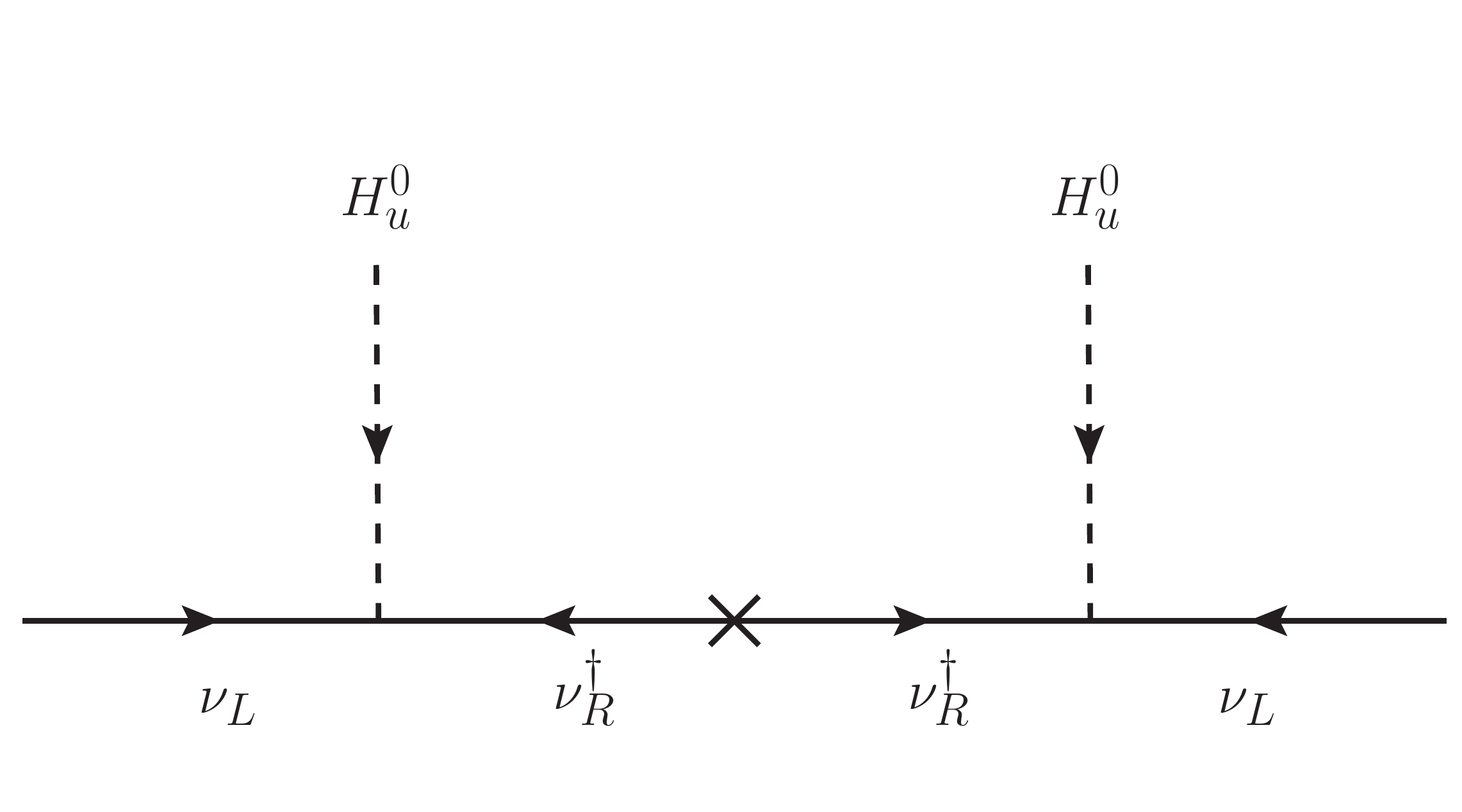}
    \end{minipage}\vspace{-.75cm}\par\vspace{\baselineskip} 
    \captionsetup{width=\textwidth, labelsep=none}
    \caption{\label{fig:seesaw}.- \small
        Feynman diagram associated with the type-I seesaw mechanism.
        The dimension-5 Weinberg operator is effectively generated when the heavy right-handed neutrinos are integrating out.}
\end{figure}

The supersymmetric type-I seesaw considers the MSSM particle content augmented with three chiral superfields, one for each generation of right-handed neutrinos (RH$\nu$).
The interactions among chiral supermultiplets are determined by the superpotential which, in this framework, contains new interactions involving RH$\nu$:
\beq \label{eq:superpot}
    W \;=\;  W_{\rm MSSM} \:+\: \nu^{c\, T}_R\, Y_\nu\, \ell_L \cdot H_u \:+\: \frac{1}{2}\, \nu^{c\, T}_R\, M_R\, \nu^c_R\,,   
\eeq
with
\bea
    W_{\rm MSSM} & = & e^{c\, T}_R\, Y_\ell\, \ell_L\cdot H_d \:+\: d^{c\, T}_R\, Y_d\,
        Q_L\cdot H_d \\
                 & + & u^{c\, T}_R\, Y_u\, Q_L\cdot H_u \:+\: \mu\, H_d\cdot H_u.
\eea
The SUSY soft-breaking interactions introduce mass terms for the superpartners as well as trilinear couplings between the new sfermions and the Higgs.
The relevant terms in our case, not including the quark sector, are:
\bea
    -{\cal L_{\rm soft}} & = & \frac{1}{2} \left( M_1 \widetilde B \widetilde B \,+\, M_2 \widetilde W \widetilde W \,+\, M_3 \tilde g \tilde g \,+\, c.c. \right) \label{eq:soft01} \\
    & + & \tilde\ell_L^\dagger\, M_{\tilde \ell}^2\, \tilde\ell_L \:+\: \tilde\e_R^{c\,T}\, M_{\tilde \e}^2\, \tilde\e_R^{c\,*} \:+\: \tilde\nu_R^{c\,T}\, M_{\tilde \nu}^2\, \tilde\nu_R^{c\,*} \\
    & + & \left( \tilde e_R^{c\, \dagger}\, A_e\, \tilde\ell_L\cdot H_d \,+\, \tilde \nu_R^{c\,\dagger}\, A_\nu\, \tilde\ell_L\cdot H_u \,+\, c.c \right) \\[2.5pt]
    & + & m_{H_d}^2\, H_u^* H_u \:+\: m_{H_d}^2\, H_d^* H_d \:+\: \dots \label{eq:soft04}
\eea
At an energy scale above the heavy RH$\nu$ mass, $\mu\geq m_{\nu^c_3}$, an effective operator with the left-handed neutrinos and Higgs fields is generated by the process in \Fig{fig:seesaw}.
Integrating out the heavy neutrinos and replacing the Higgs by its vev at low energies, the following mass matrix is produced:
\beq \label{eq:seesaw1}
    M_\nu \:=\: -\frac{\vu^2}{2}\; Y_\nu^T\, M_R^{-1}\, Y_\nu,
\eeq
where $\vu=\upsilon_h\, \sin\beta$ and $\upsilon_h=246\,{\rm GeV}$. %
Neutrino oscillations are directly related to $M_\nu$ in \Eq{eq:seesaw1} and provide information about the squared-mass differences and mixing of neutrinos, see \Tab{tab:nufit}.
However, those measurements are not enough to fully reconstruct the UV parameters of the model, namely $Y_\nu$ and $M_R$.

Extensions of the SM may provide additional information through related observables, like CLFV processes. One of the best examples is supersymmetric extensions of the SM supplemented with a type-I seesaw, where slepton soft-breaking masses are related to the neutrino Yukawa couplings.
However, additional assumptions are usually made to simplify the analysis of the phenomenology of these models.
The minimal strategy consists in presuming universal soft-masses and a concrete structure for the neutrino Yukawa couplings at the GUT scale.
Then non-universal entries are generated in the soft masses through the renormalization group evolution (RGE), proportional to the hypothesised Yukawa couplings.
It is important to remark that these RGE contributions are always present in SUSY models irrespective of the presence of additional non-universal entries at the GUT scale.
Therefore, barring accidental cancellations, these effects are the minimal outcome of supersymmetric seesaw models.
\begin{table}
	\centering
	{\renewcommand{\arraystretch}{1.25}
    \resizebox{0.65\columnwidth}{!}{
	\begin{tabular}{c c c}
    \toprule
    \bf Observable & \bf Normal Hierarchy & \bf Inverted Hierarchy \\[2.5pt]
    \midrule
    $\theta_{12}~ (º)$ & $33.82^{+0.78}_{-0.76}$ & $33.82^{+0.78}_{-0.76}$ \\[5pt]
    $\theta_{23}~ (º)$ & $48.3^{+1.1}_{-1.9}$ & $48.6^{+1.1}_{-1.5}$\\[5pt]
    $\theta_{13}~ (º)$ & $8.61^{+0.13}_{-0.13}$ & $8.65^{+0.13}_{-0.12}$ \\[5pt]
    $\Delta m_{12}^2~ (10^{-5}\, \eV^2)$ & $7.39^{+0.21}_{-0.20}$ & $7.39^{+0.21}_{-0.20}$ \\[5pt]
    $\Delta m_{3\ell}^2~ (10^{-3}\, \eV^2)$ & $2.523^{+0.032}_{-0.030}$ & $-2.509^{+0.032}_{-0.030}$ \\[2.5pt]
    \bottomrule
    \end{tabular}}}
\captionsetup{width=0.8\textwidth,labelsep=none}
\caption{\label{tab:nufit} \small
        .- Global fit on neutrino observables by \href{http://www.nu-fit.org}{NuFIT 4.1. (2019)} \cite{Esteban:2018azc}.
        Similar results have been found in \cite{Ref_NeutOthers}.}
\end{table}

In this framework, we perform the analysis in two {\it limit} cases: one in which the rotation angles in $V_L^\nu$ are small, or CKM-like, and another where the mixing is large, or PMNS-like.
These two scenarios should span any intermediate possibility so that general conclusions can be fairly derived.
For instance, based on an underlying $SO(10)$ gauge symmetry, one may consider that $Y_\nu$ and $Y_u$, in the basis of diagonal Yukawas for charged leptons and down-type quarks, are deeply connected at the unification scale through relations such as \cite{Masiero:2002jn, Calibbi:2006nq, Calibbi:2011dn, Calibbi:2012gr}:
\begin{flalign}
    & \text{- Small Mixing (CKM-like): } \yckm = \kGJ\, Y_u. & \label{eq:ckmlike_a} \\[5pt]
    & \text{- Large Mixing (PMNS-like): } \ypmns = \kGJ\, Y_u^{\rm diag}\, \vpmns^T & \label{eq:pmnslike_a}
\end{flalign}

where, in these equations, we have introduced a Georgi-Jarlskog (GJ) factor, $\kGJ$, which may arise in some GUT realisations due to the different representations of the unified group that mix to generate the SM Higgs doublet.

Within a $SO(10)$ model, for example, if the dominant contribution to the Yukawa interactions is due to a Higgs field transforming as a $\bm{10}$, a perfect unification between neutrinos (charged leptons) and up-type (down-type) quarks is expected, so $\kGJ=1$.
Conversely, if the dominant contribution comes from a $\bm{126}$ representation, a factor $\kGJ=3$ appears between leptons and quarks.
Another possibility is having an effective Higgs field transforming as a $\bm{120}$, which may be the product of a $\bm{45}$ and ${\bf 10}$ representations.
The $\bm{45}$ can acquire a nonzero vev in the flat direction $B-L+\kappa T_3$, which preserves the SM and distinguishes between RH fermions\footnote{$T_3$ refers to the third component of a $SU(2)_R$ gauge group which is spontaneously broken afterwards.} \cite{Ref_GJflavour}.
In fact, complete flavour models usually require the combined effect of more than one representation to generate dissimilar hierarchies among generations \cite{Ref_ourGJflavour}.
For instance, unification in the down sector as in the classical Georgi-Jarslkog scheme demands $|y_\tau/y_b|_{\rm GUT} = 1$ and $|y_\mu/y_s|_{\rm GUT} = \kGJ =3$.
However, nowadays, these relations are no longer favoured phenomenologically \cite{Antusch:2008tf, Antusch:2009gu, Antusch:2011xz, Antusch:2013rxa} but the updated range
\beq \label{eq:ymuys_pheno}
    \big| y_\mu/y_s \big|_{\rm GUT} = [\,2.5,\; 6.5\,],
\eeq
obtained in \cite{Antusch:2011xz}.
This is the reason we take to generalize our $\kGJ$ factor to $\kGJ = B-L+\kappa T_3$.
In the case of neutrino Yukawa couplings, we are allowed to consider that the dominant contribution to the up and neutrino Yukawas comes from the representation giving rise to this $\kGJ$ factor.
Then, varying $\kappa$ in the interval where \Eq{eq:ymuys_pheno} is satisfied, one observes that
\beq \label{eq:Ynu120}
    \big| Y_\nu \big|\:=\: [\,0,\; 1/2\,]\; \big|Y_u\big|.
\eeq
This is the interval taken for the $\kGJ$ factor in our numerical analysis.

We explore the parameter space of seesaw NUHM3 models in which some of the stringent conditions of the typical mSUGRA models are relaxed.
We introduce three additional degrees of freedom in the scalar soft-breaking sector: instead of one common scalar mass, we will consider the following four
\beq \label{eqn:NUHM3}
    m_0^{(1,2)} \neq m_0^{(3)} \neq m_{H_u} \neq m_{H_d}\,
\eeq
where we have included a 
non-universal, but diagonal, charged-slepton mass matrix in the basis of diagonal charged-lepton Yukawa couplings at the GUT scale.
The supersymmetric sector of the models is determined by five parameters at the GUT scale,
\beq
    m_0^{(1,2)}, \hspace{0.5cm} m_0^{(3)}, \hspace{0.5cm} M_{1/2}, \hspace{0.5 cm} \tan\beta, \hspace{0.5cm} A_0,
\eeq
and two more at the EW scale,
\beq
    \mu, \hspace{1.cm} M_{A^0},
\eeq
which can be taken in exchange of $ m_{H_u}$ and $m_{H_d}$,

\section{RGEs and lepton flavour violation}
\label{sec:RGE}
The introduction of RH neutrinos makes the effect of the RGEs specially relevant for sleptons \cite{Ref_SUSY_RHn, Hisano:1995cp}.
During the running, the heavy neutrinos induce off-diagonal entries in the slepton soft terms through radiative corrections.
Those flavour-violating interactions allow for CLFV processes that otherwise, within the SM, would be greatly suppressed by the neutrino masses.
The main effect occurs for the soft-mass matrices of the LH sleptons and can be worked out by solving the RGEs
{\small \bea \label{eq:REG}
    \mu \frac{d}{d\mu}\bigg(\MsL\bigg) & = & \mu\frac{d}{d\mu}\bigg( \MsL \bigg)_{\rm MSSM} \\
    & + & \frac{1}{16\pi^2} \bigg( \MsL Y_\nu^\dagger Y_\nu \:+\: 
        Y_\nu^\dagger Y_\nu\, \MsL \:+\: 2\, \left(Y_\nu^\dagger 
        M^2_{\tilde \nu} Y_\nu \:+\: m_{H_u}^2 Y_\nu^\dagger Y_\nu \:+\: 
        A_\nu^\dagger A_\nu\right) \bigg),\qquad
\eea}
where the first term denotes the MSSM contribution in the absence of \RHn,
{\small \bea
    \mu\frac{d}{d\mu}\bigg( \MsL \bigg)_{\rm MSSM} & = & \frac{1}{16\pi^2} \bigg( \MsL 
        Y_\ell^\dagger Y_\ell \,+\, Y_\ell^\dagger Y_\ell \MsL \:+\: 2\left(Y_\ell^\dagger M_{\tilde \e}^2 Y_\ell \,+\, m_{H_d}^2 Y_\ell^\dagger Y_\ell \,+\, A_\ell^\dagger A_\ell \right) \bigg)\qquad \\
    & - & \mathbb{1}\, \left( \frac{6}{5}g_1^2 \left|M_1\right|^2 \,+\, 6\, g_2^2 \left|M_2\right|^2
        \right) \:+\: \mathbb{1}\, \frac{3}{5} g_1^2 S
\eea}
with $S\equiv {\rm Tr}[M_{\tilde Q_L}^2+M_{\tilde d}^2-2M_{\tilde u}^2-\MsL+M_{\tilde \e}^2]-m_{H_d}^2+m_{H_u}^2$.
In the basis of diagonal RH$\nu$ and charged-lepton Yukawas, the leading log approximation is proportional to the square of the neutrino Yukawas as:
\beq \label{eqn:RGEsoftmass} 
    \left(M_{\tilde \ell}^2\right)_{i\neq j} ~\simeq~ -\frac{2\, m_0^2 + m_{H_u}^2 + A_0^2}{16\, \pi^2}\; \sum_k Y_{\nu, ki}^* Y_{\nu, kj}\, \log \left( \cfrac{\mgut^2}{m_{N_k}^2}\right),
\eeq
where we take the limit $m_0^{(1,2)} \simeq m_0^{(3)} \simeq m_0$ and approximate $\mgut$ to be of the order of the scale at which the soft terms appear in the Lagrangian (the typical scale of SUSY-breaking transmission).
The main effect, with hierarchical Yukawas, is due to the heaviest Majorana neutrino and happens before its decoupling at $\mu > m_{\nu^c_3}$.
Trilinear couplings receive similar corrections, although they have a smaller impact on the CLFV observables studied here.
In contrast, no flavour violation is produced in the RH charged-slepton sector at one-loop, since the RGEs only depend on $Y_\ell$ and on the gauge couplings, hence they are diagonal in the basis where $Y_\ell$ is diagonal.
The off-diagonal elements produced radiatively enter the total $6\times 6$ slepton mass matrix as small insertions (compared to the diagonal terms) in the LL and LR/RL sector:
\beq \label{eq:Msl}
    M_{\tilde \ell}^2 ~=~ \begin{pmatrix}
                            \Delta_{LL} & \Delta_{LR} \\
                            \Delta_{LR}^\dagger & \Delta_{RR}
                            \end{pmatrix}
\eeq
\bea
    \Delta_{LL} & = & M_{\tilde \ell_L}^2 \:+\: \frac{v_d^2}{2}\, Y_\ell^\dagger Y_\ell 
        \:+\: \mathbb{1}~ m_Z^2\, \cos2\beta\, \left(-\frac{1}{2} + \sin^2\theta_w 
        \right) \\
    \Delta_{RR} & = & M_{\tilde \e_R}^2 \:+\: \frac{v_d^2}{2}\, Y_\ell^\dagger Y_\ell 
        \:-\: \mathbb{1}~ m_Z^2\, \cos2\beta\, \sin^2\theta_w\\
    \Delta_{LR} & = & \frac{v_d}{\sqrt{2}} \left( A_\ell \:-\: \mu^*\, Y_\ell\, \tan\beta
        \right).
\eea

\section{Observables}
\label{sec:obs}
Up to now, we have defined the supersymmetric model that we analyse in this project and its RGE evolution to the electroweak scale.
The next step will be to compare its predictions with the low-energy observables, to constrain the allowed parameter space or to find possible discrepancies from the SM predictions. 

The first observable we have to reproduce is the recently measured value of the Higgs mass, which is a strong constraint on any supersymmetric extension of the SM.
Then, as we are mainly interested in the leptonic sector, we concentrate on two main observables: the anomalous magnetic moment of the muon and the CLFV process $\mu \to e \gamma$.  

\subsection{Higgs mass}
\label{subsec:Higgs}
Previous works have extensively discussed how to accommodate the $125\, \GeV$ observed Higgs boson \cite{Ref_LHCHIggs} within a minimal supersymmetric framework %
\cite{Arbey:2011ab, Cao:2012fz, Carena:2013ytb, Ref_ourHiggs}. %
In the MSSM, it is known that the tree-level value of the lightest Higgs mass is bounded from above by $M_Z$ whilst radiative corrections, coming from the fermion-sfermion loops, may increase it up to $135~\GeV$ \cite{Ref_SusyHiggsUppBound}.
As those corrections are proportional to the corresponding fermion Yukawa couplings, the dominant contribution is due to the top-stop diagram and can be written as
\beq
    \Delta m_h ~\simeq~ \frac{3}{4\pi^2}\, \cos^2\alpha\, y_t^2\, m_t^2\, \left[\, 
        \ln\left(\frac{m_{\tilde t_1}m_{\tilde t_2}}{m_ t^2} \right) \,+\, \Delta_{\rm thr} \right],
\eeq
where $\alpha$ is the mixing angle between the scalar components of $H^0_u$ and $H^0_d$ after EWSB and $\Delta_{\rm thr}$ stands for the threshold corrections dependent on the stop mixig \cite{Martin:1997ns}.
While constrained versions of the MSSM, such as mSUGRA, GMSB or AMSB, usually have difficulties to generate the observed mass, scenarios with non-universal conditions at the GUT scale are able to improve their predictions and provide realisations with a low amount of fine-tuning \cite{Arbey:2011ab, Ref_RNS125Higgs}.

In the models considered here, the stop mass is determined by the scalar mass parameter $m_0^{(3)}$ at the GUT scale.
As we will see below, in order to obtain the adequate mass, large values for $m_0^{(3)}$ are expected,
\beq
    m_0^{(3)} \gtrsim 4\, \TeV,
\eeq
with stops masses in the few-TeV regime,
\beq
    m_{\tilde t_1},\, m_{\tilde t_2} \gtrsim 2.5\, \TeV.
\eeq

\subsection{Anomalous magnetic moment of the muon}
\label{subsec:g2}
\begin{figure}[t!]
  \centering
    \begin{minipage}[c]{\textwidth}
        \centering
        \includegraphics[width=0.475\textwidth]{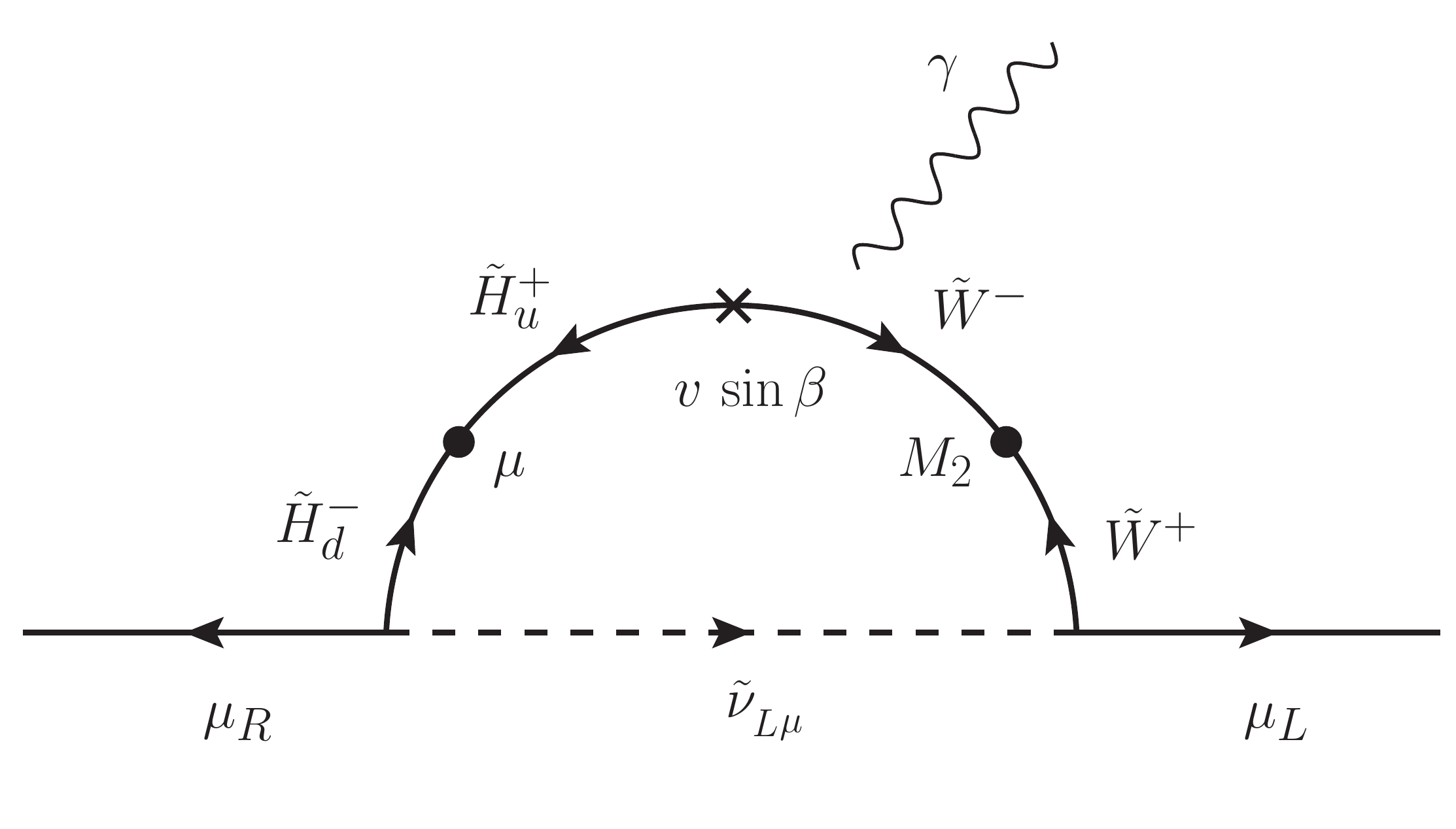}
        \hspace{0.cm}
        \includegraphics[width=0.475\textwidth]{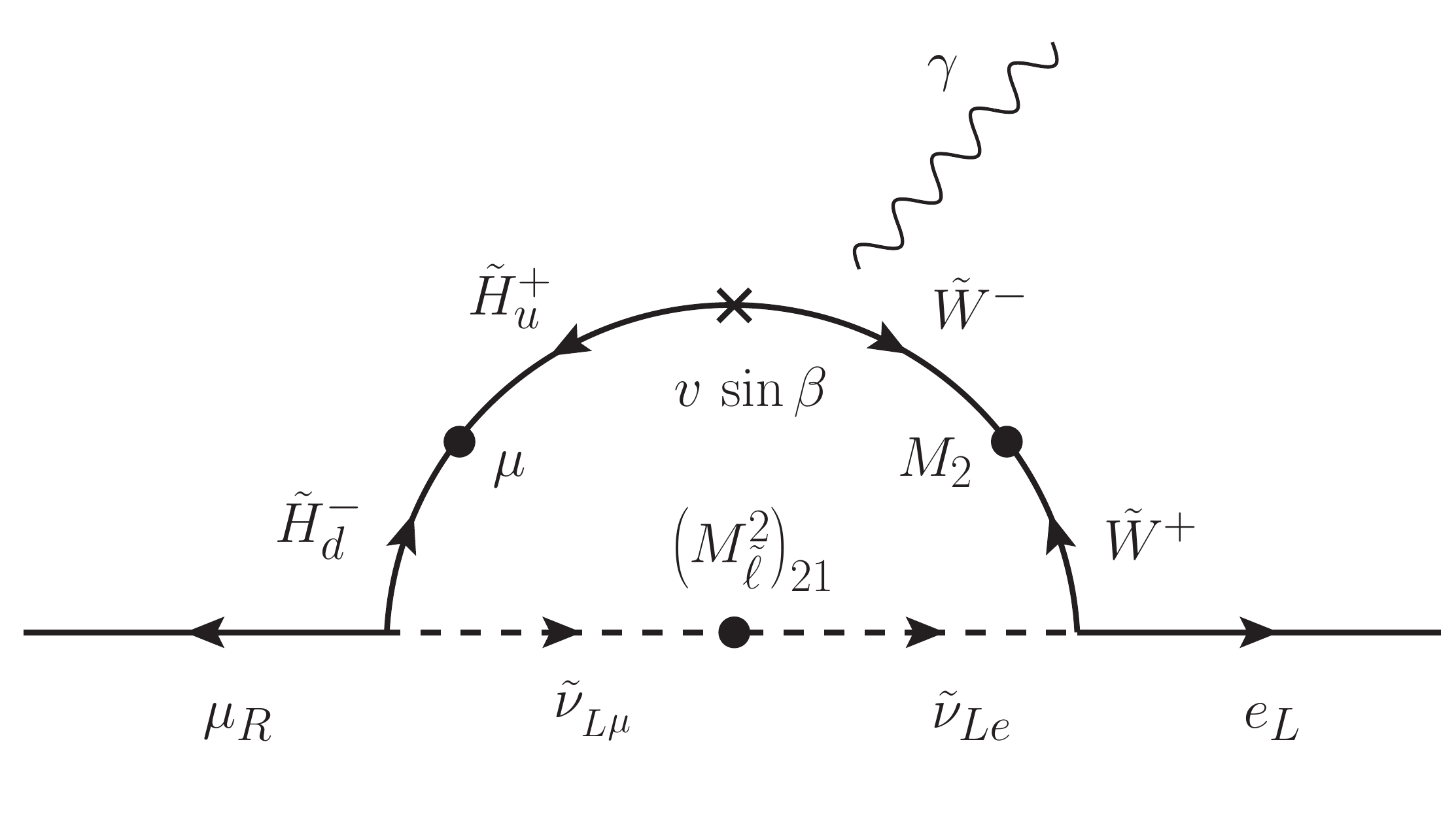}
    \end{minipage}\vspace{-.75cm}\par\vspace{\baselineskip} 
    \captionsetup{width=\textwidth, labelsep=none}
    \caption{\label{fig:g2_mey_dom}.- \small
        Diagrammatic representation of the leading contribution to $\gmususy$ (left) and ${\BR}(\mu\to\e\gamma)$ (right) in our models.}
\end{figure}
In the MSSM, leptons receive supersymmetric corrections to their anomalous magnetic moment due to neutralino and chargino loops that effectively generate the dipole operators, defined  in \Eq{appeq:dipole_op} in Appendix \ref{appsubsec:mey_g2_MIA}, where we can find the full expressions, \cite{Hisano:1995cp, Moroi:1995yh},
\beq
    a_\ell^{\rm susy} ~=~ a_\ell^{(c)} \;+\; a_\ell^{(n)}.
\eeq
The supersymmetric amplitude is usually dominated by the processes where the chirality flip of the fermion occurs at the vertex, which is proportional to the Yukawa coupling and therefore $\tan\beta$-enhanced.
The mass insertion approximation (MIA) \cite{Hall:1985dx, Gabbiani:1996hi, Misiak:1997ei,Dedes:2015twa} allows us to see this explicitly by means of expanding the full amplitude, extracting the relevant diagrams and identifying the main parameters.
It has been implemented in Appendix \ref{appsubsec:mey_g2_MIA}.
The diagramatic interpretation of the dominant processes is depicted in \Fig{fig:g2_mey_dom} (left).

\begin{figure}[t!]
  \centering
    \begin{minipage}[c]{\textwidth}
        \centering
        \includegraphics[width=0.475\textwidth]{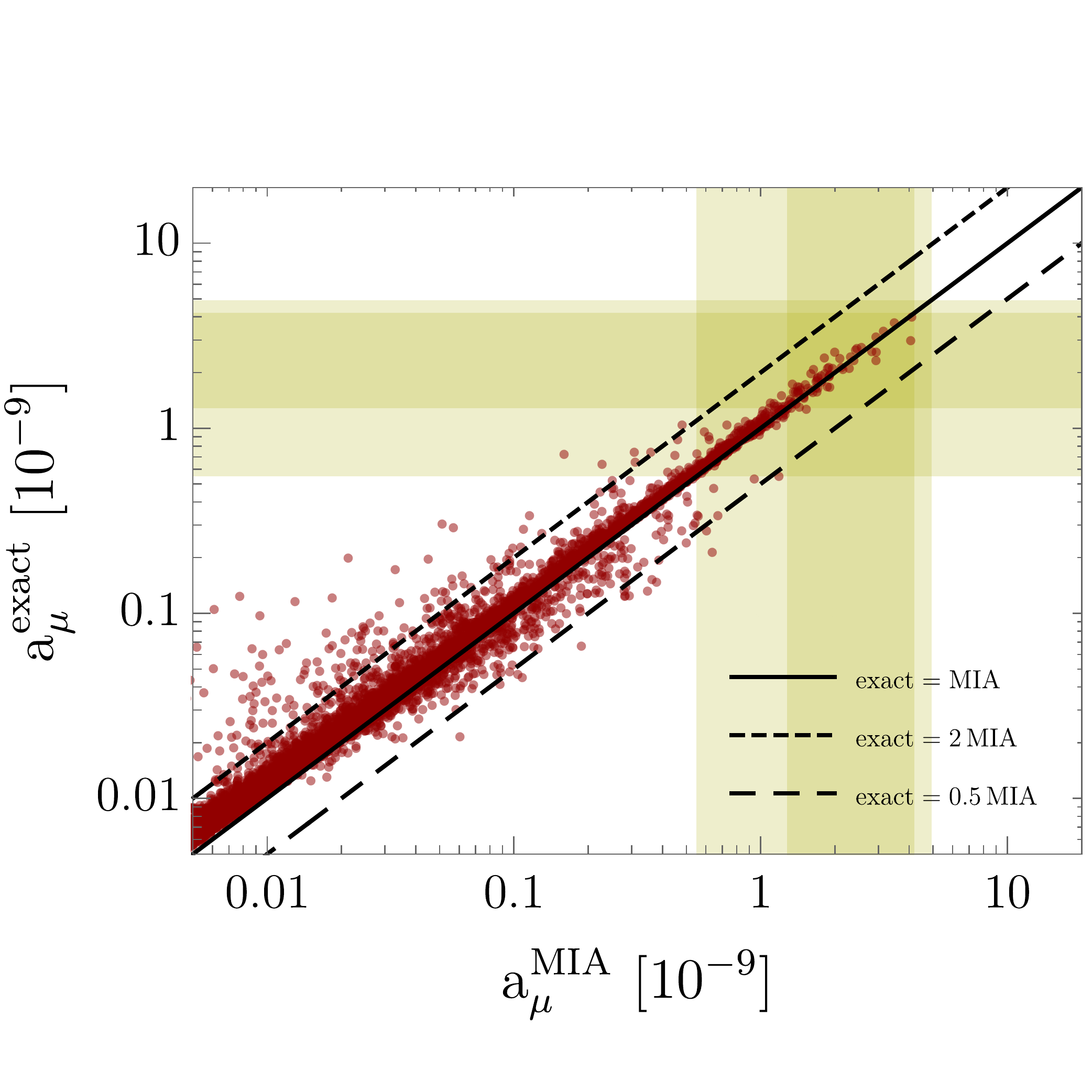}
        \hspace{0.5cm}
        \includegraphics[width=0.475\textwidth]{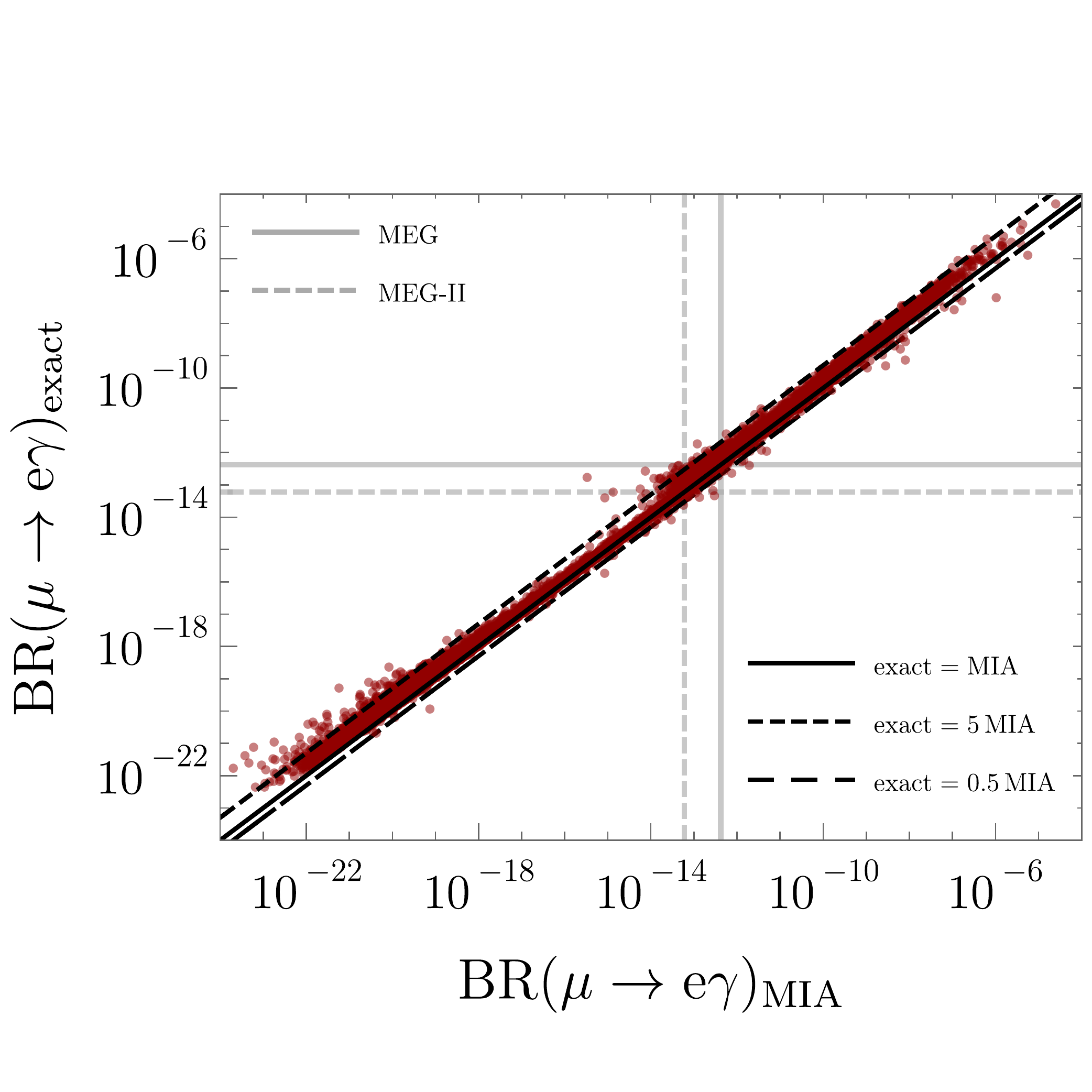}
    \end{minipage}\vspace{-1.cm}\par\vspace{\baselineskip} 
    \captionsetup{width=\textwidth, labelsep=none}
    \caption{\label{fig:amu_mey_comp} \small
        .- Comparison between the exact values for $a_\mu^{\rm susy}$ (left) and ${\rm BR}(\mu\to\e\gamma)$ (right) computed by \texttt{SPheno} versus the results obtained through the mass insertion approximation in Appendix \ref{appsubsec:mey_g2_MIA}.}
\end{figure}

In the light higgsino-world scenario, the LSP is the neutralino, which is mainly higgsino and quasi-degenerate in mass with the second-lightest neutralino (NLSP) and the lightest chargino.
We observe that the region where $\gmususy$ is within the $3\sigma$ range exhibits the following hierarchies between masses: $\mu \ll M_2 \lesssim m_{\tilde \ell_L}$.
Then, the process is expected to be dominated by the chargino loop\footnote{More details about the derivation of the chargino and neutralino dominant terms under the MIA can be found in Appendix \ref{appsubsec:mey_g2_MIA}.}, since its loop function for $x=\mu^2/m_{\tilde \ell_L}^2\ll 1$ is the largest one (see \Fig{appfig:loop}).
Therefore,
\beq \label{eq:g2MIA}
        a_\mu^{\rm susy} ~\simeq~ -\frac{\alpha_2}{4\pi}\, \frac{m_\mu^2}{m_{\tilde
        \nu_\mu}^2}\, \frac{M_2\, \mu}{M_2^2 \,-\, \mu^2}\, 
        F_2^{c}(x_{2\tilde{\nu}_\mu},x_{\mu\tilde{\nu}_\mu})\, \tan\beta,
\eeq
where $x_{2{\tilde\nu}_\mu}=M_2^2/m_{\tilde \nu_\mu}^2$, $x_{\mu{\tilde\nu}_\mu}=\mu^2/m_{\tilde \nu_\mu}^2$ and $F_2^c(x_1,x_2)\equiv f_2^{(c)}(x_1)-f^{(c)}_2(x_2)$ with $f_2^c(x)$ the loop function provided in Appendix \ref{appsec:loop}.
We compare the exact result worked out by \texttt{SPheno-4.0.4} \cite{Porod:2003um, Porod:2011nf} versus the MIA expression in \Fig{fig:amu_mey_comp} (left) and notice that the second works quite well for most of the points.
Some deviations appear for isolated points in the region where $a_\mu^{\rm susy}$ is very small and contributions from other diagrams may compete and become important. 

From \Eq{eq:g2MIA}, another phenomenological consequence can be inferred: a SUSY contribution that accounts for the current discrepancy between the experimental and the SM theoretical value of the muon anomalous magnetic moment will require light sneutrinos in the second generation.
In the models analysed here, the masses for sfermions of the first two generations are determined by the scalar mass $m_0^{(1,2)}$ at the GUT scale. 
Therefore, contrary to $m_0^{(3)}$ (see discussion in Section \ref{subsec:Higgs}), we expect quite small values for $m_0^{(1,2)}$ to reproduce $\Delta a_\mu$,%
\beq \label{eq:decoup_spect}
    m_0^{(1,2)} \ll m_0^{(3)}.
\eeq
In practice, \Eq{eq:decoup_spect} leads to a decoupled spectrum for sfermions where the third generation is significantly heavier than the first two ones.
\subsection{CLFV observables}
\label{subsec:CLFV}
CLFV transitions such as $\ell_j \to \ell_i\gamma$, $\ell_j \to 3\ell_i$ and $\mu\to\e$ {\it in Nuclei} are extremely rare in the SM.
Actually, we do {\it not} expect observing them if the SM is only theory at low energies.
The experimental limits on these processes are collected in \Tab{tab:LFV}.

In the MSSM, the $\ell_j\to\ell_i$ transitions are dominated by the dipole operators arising at one-loop level by the exchange of charginos-sneutrinos and neutralinos-charged sleptons, \Eq{appeq:dipole_op}.
The flavour switch is feasible due to the flavour-violating interactions generally present in the soft-breaking terms (trilinears and soft masses).
Actually, the diagrams for $\ell_j \to \ell_i$ and $a_\ell^{\rm susy}$ are essentially identical, except for the flavour transition happening in the former which is absent in the latter.
Motivated by this, several articles have evaluated the correlations between $\mu\to e\gamma$ and $(g\text{-}2)_\mu$ in the MSSM \cite{Graesser:2001ec, Chacko:2001xd, Bi:2002ra, Isidori:2007jw, Ibrahim:2015hva, Kersten:2014xaa, Lindner:2016bgg}, although a dedicated discussion in the context of the light higgsino-world scenario is still lacking.

As discussed in Section \ref{subsec:g2}, in the models under consideration, the lightest neutralinos and chargino are higgsino-like and quasi-degenerate in mass.
Assuming that the off-diagonal entries are much smaller than the diagonal ones and applying the MIA for sneutrino and chargino propagators, we find that the branching fraction given by
\beq \label{eq:BRmey}
    \frac{{\rm BR}(\mu\to \e\gamma)}{{\rm BR}(\mu\to \e \nu_\mu \bar\nu_\e)} ~=~ \frac{48\pi^3\alpha}{G_F^2}\, \left(\,|a_L^{\mu\e}|^2 \,+\, |a_R^{\mu\e}|^2 \,\right)
\eeq
is well reproduced by the following amplitudes
\bea
    a_R^{\mu\e} & \simeq & \frac{\alpha_2}{4\pi}\; 
        \frac{\left(M_{\tilde \ell_L}^2\right)_{21}}{m_{\tilde \nu}^4}\; \frac{M_2\,
        \mu}{M_2^2 \,-\, \mu^2}\, F_3^{c}(x_{2\tilde{\nu}},x_{\mu\tilde{\nu}})\, \tan\beta, \label{eq:mey_AL_MIA} \\
    a_L^{\mu\e} & \simeq & 0, \label{eq:mey_AR_MIA}
\eea
where $F_3^c(x_2,x_\mu)\equiv f_3^{(c)}(x_2)-f_3^{(c)}(x_\mu)$ and $f_3^{(c)}(x)$ the loop function in Appendix \ref{appsec:loop}.
The diagramatic interpretation of \Eq{eq:mey_AL_MIA} is very similar to the one for $a_\mu^{\rm susy}$ but with a flavour-changing insertion inside the loop, see \Fig{fig:g2_mey_dom} (right).
The accuracy of the approximation is compared to the exact value in \Fig{fig:amu_mey_comp} (right panel).
A good agreement is found between both results.

Finally, it is worth commenting the processes $\mu\to 3\e$ and $\mu-\e$ conversion.
In this case, both receive contributions from penguin diagrams (with Z-boson and photon exchanges), as well as, from box-diagrams.
However, again, the $\gamma$-penguin transition (directly connected to $\mu\to\e\gamma$), being  $\tan\beta$-enhanced and not suppressed by $M_Z$, dominates and the following simple relations hold:
\bea
    {\BR(\mu\to3e)} & \simeq & \frac{\alpha}{3\pi}\, \left(\log 
        \frac{m_\mu^2}{m_e^2} - 3 \right)\, \times\, \BR(\mu \to e \gamma), 
        \label{eq:li3lj} \\[5pt]
    {{\rm CR}(\mu\, {\rm N} \to \e\, {\rm N})} & \simeq & \alpha\, \times\, 
        \BR(\mu\to\e\gamma). \label{eq:mu-e_conv}
\eea
\begin{table}[t]
	\centering
	{\renewcommand{\arraystretch}{1.5}
	\resizebox{0.75\columnwidth}{!}{
	\begin{tabular}{c c c}
    \toprule
    \bf LFV Process & \bf Current Limit & \bf Future Limit \\[2.pt]
    \midrule
        BR($\mu\to\e\gamma$) & $4.2\times10^{-13}$ (\texttt{MEG at PSI}\cite{TheMEG:2016wtm}) & $6\times10^{-14}$ \texttt{(MEG\,II} \cite{Baldini:2018nnn}) \\
        BR($\mu\to3\e$)     & $1.0\times10^{-12}$ (\texttt{SINDRUM}\cite{Bellgardt:1987du}) & ~~\quad $10^{-16}$ (\texttt{Mu3e}\cite{Blondel:2013ia}) \\
        CR$(\mu-\e)_{A_l}$  & $-$ & ~~\quad $10^{-17}$ (\texttt{Mu2e}\cite{Bartoszek:2014mya}, \texttt{COMET}\cite{Blondel:2013ia}) \\
        BR$(\tau\to\e\gamma)$  & $3.3\times 10^{-8}$ (\texttt{BaBar}\cite{Aubert:2009ag}) & \qquad $5\times10^{-9}$ (\texttt{Belle\,II}\cite{Aushev:2010bq}) \\
        BR$(\tau\to\mu\gamma)$ & $4.4\times 10^{-8}$ (\texttt{BaBar}\cite{Aubert:2009ag}) & \qquad $10^{-9}$ (\texttt{Belle\,II}\cite{Aushev:2010bq}) \\
        BR$(\tau \to 3\e)$  & $2.7\times10^{-8}$ (\texttt{Belle}\cite{Miyazaki:2011xe}) & \qquad $5\times10^{-10}$ (\texttt{Belle\,II}\cite{Aushev:2010bq}) \\
        BR$(\tau\to3\mu)$   & $2.1\times10^{-8}$ (\texttt{Belle}\cite{Miyazaki:2011xe}) & \qquad $5\times10^{-10}$ (\texttt{Belle\,II}\cite{Aushev:2010bq}) \\
    \bottomrule
    \end{tabular}}}
\captionsetup{width=\textwidth,labelsep=none} \small
\caption{\label{tab:LFV}.- 
        Current and future expected limits on CLFV processes.}
\end{table}
\subsection{\boldmath $\gmu$ vs. $\mu\to\e\gamma$}
\label{subsec:g2vsmey}
The parallelism between the amplitude in \Eq{eq:g2MIA} and the one in \Eq{eq:mey_AL_MIA} is evident.
Taking the ratio between \Eq{eq:mey_AL_MIA} and \eqref{eq:g2MIA}, it is obtained
\beq \label{eq:ratio}
   \frac{a_R^{\mu\e}}{a_\mu^{\rm susy}} ~\simeq~ \frac{\left(M_{\tilde 
        \ell_L}^2\right)_{21}}{m_\mu^2}\, 
        \frac{F_3^c(x_{2\tilde{\nu}},x_{\mu\tilde{\nu}})} 
        {F_2^c(x_{2\tilde{\nu}_\mu},x_{\mu\tilde{\nu}_\mu})}.
\eeq
The branching ratio in \Eq{eq:BRmey} can therefore be rewritten as:
\bea 
    \frac{{\rm BR}(\mu\to \e\gamma)}{{\rm BR}(\mu\to \e \nu_\mu \bar\nu_\e)} 
    & \simeq & \frac{48\pi^3\alpha}{G_F^2\, m_\mu^4}\, \bigg|a_\mu^{\rm susy}\, 
        \frac{\left(M_{\tilde \ell_L}^2\right)_{21}}{m_{\tilde \nu}^2}\,
        \frac{F_3^c(x_{2\tilde{\nu}},x_{\mu\tilde{\nu}})} 
        {F_2^c(x_{2\tilde{\nu}_\mu},x_{\mu\tilde{\nu}_\mu})} \,\bigg|^2 
        \label{eq:ratio_analytic} \\
    & \approx & \big[0.06,\, 4\big]\, \left|\frac{a_\mu^{\rm susy}}{10^{-9}}\,
        \frac{\left(M_{\tilde \ell_L}^2/m_{\tilde \nu}^2\right)_{21}}{10^{-4}} \,\right|^2 \times 10^{-12}, \label{eq:ratio_num}
\eea
where the interval is due to the variation of the loop functions in the obtained ranges $x_{\mu\tilde \nu}\in [0.01,5]$ and $x_{2\tilde \nu}\in [0.1,100]$, see \Fig{appfig:loop}.
From \Eqs{eq:ratio_analytic}{eq:ratio_num}, one may see that, for those values of $M_2,\, \mu \text{ and } m_{\tilde{\nu}_\mu}$ where $a_\mu^{\rm susy}$ is within the $3\sigma$ range, an off-diagonal element in the slepton soft-mass matrix around $10^{-4}$ is required to fulfill current and future limits on ${\rm BR}(\mu\to\e\gamma$). %

\begin{figure}[t!]
  \centering
    \begin{minipage}[c]{\textwidth}
        \centering
        \includegraphics[width=0.475\textwidth]{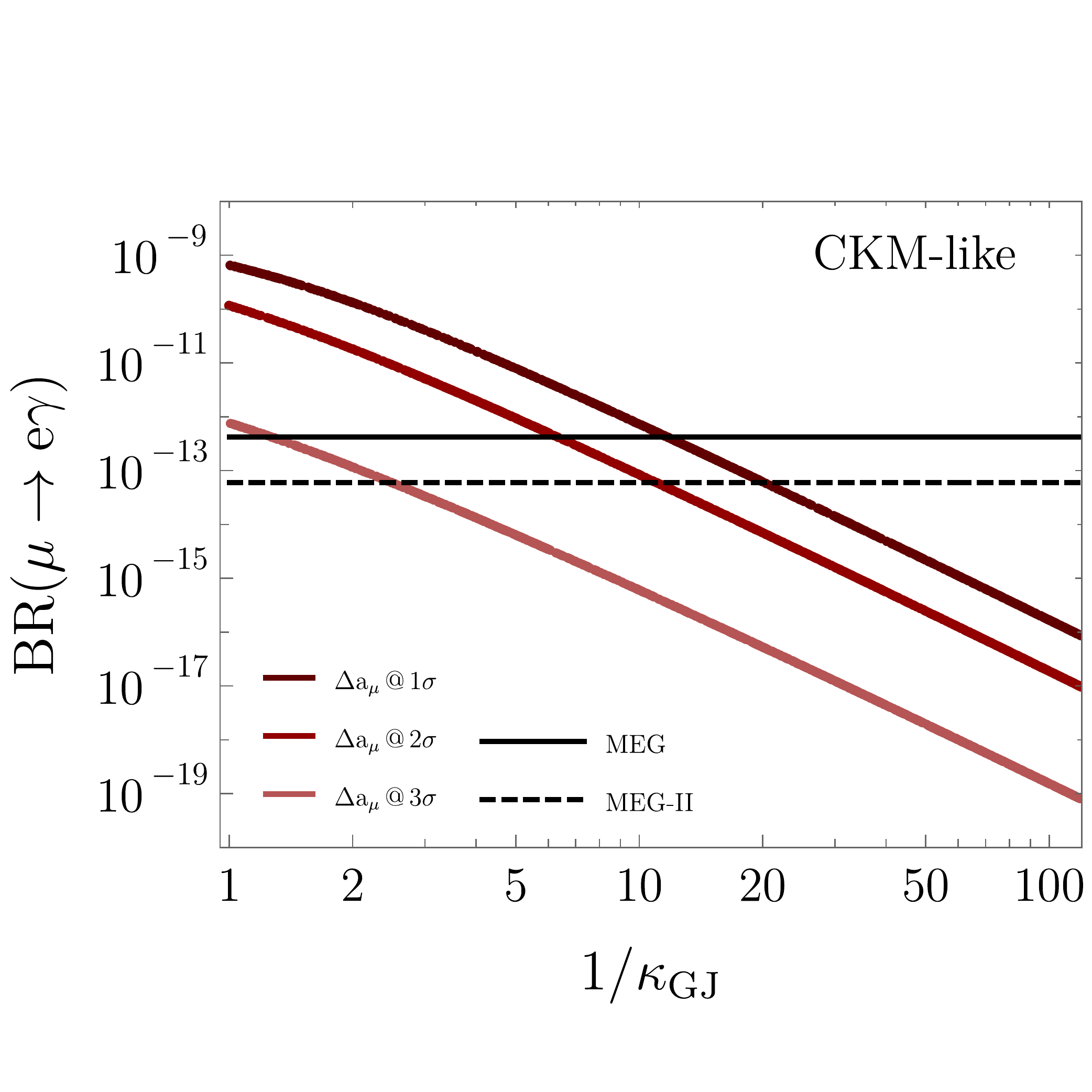}
        \hspace{0.5cm}
        \includegraphics[width=0.475\textwidth]{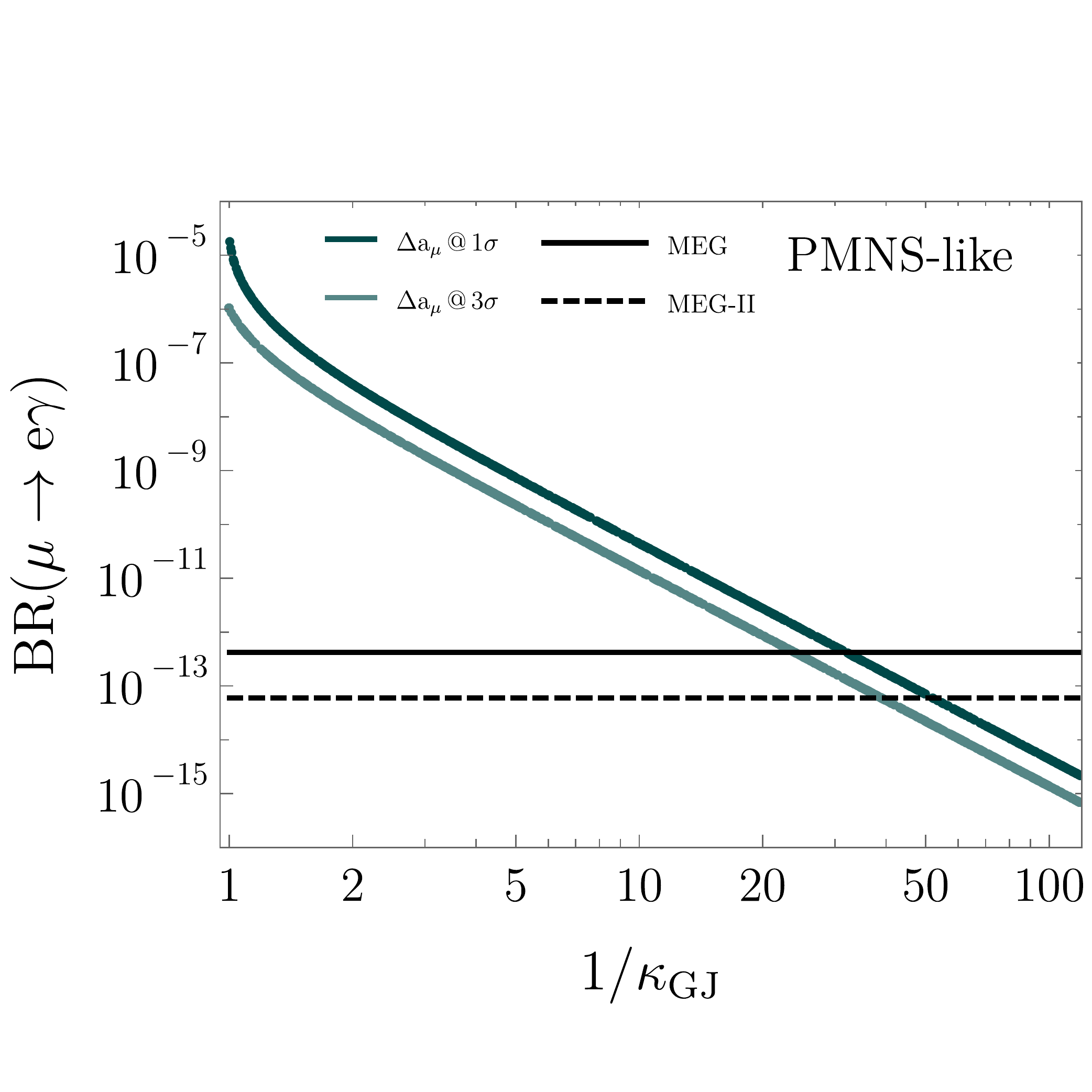}
    \end{minipage}
    \vspace{-1.cm}\par\vspace{\baselineskip}
    \captionsetup{width=\textwidth,labelsep=none}
    \caption{\label{fig:mey_gmu}.- \small
    In each case, we choose some representatives points that satisfy all the 
    constraints, and reproduce $\Delta a_\mu$ at different levels of accuracy, to test how ${\rm BR}(\mu\to\e\gamma)$ varies with $\kGJ$.
    The black continuous (dashed) line signals the current (future) limit from \texttt{MEG 
    (MEG-II)} \cite{TheMEG:2016wtm, Baldini:2018nnn}.
    }
\end{figure}

Additionally, \Eq{eqn:RGEsoftmass} reveals that the induced off-diagonal term is proportional to the square of the GJ factor introduced in \Eqs{eq:ckmlike_a}{eq:pmnslike_a}.
Therefore, the absence of any signal in the FC process becomes an indirect upper bound on $\kGJ$ for our benchmark scenarios.
We have explicitly checked the relation between ${\rm BR}(\mu\to\e\gamma)$ and $\kGJ$ for some representative points.
The result is displayed in \Fig{fig:mey_gmu}.
We select some points that, among other constraints, fulfill $\Delta a_\mu$ at different levels of accuracy when $\kGJ=1$ and evaluate how ${\rm BR}(\mu\to\e\gamma)$ is modified when $\kGJ$ decreases.
As expected, ${\rm BR}(\mu\to\e\gamma)$ is reduced.
Consequently, a factor
\beq \label{eq:kGJ_from_mey}
    \kGJ \simeq 1/3~ \text{(CKM-like)} \hspace{1.5cm} \kGJ \simeq 1/50~ \text{(PMNS-like)}
\eeq
would be needed to conciliate both observables, considering future sensibility from \texttt{MEG-II}.
Notice that \Eq{eq:kGJ_from_mey} should be taken as a guiding factor, since the RGE effects are not linear in the neutrino Yukawa couplings and the implications on $\gmususy$ may change for different values of $\kGJ$.
Then, $kGJ$ in \Eq{eq:kGJ_from_mey} can change to some extent in the numerical analysis of Section \ref{sec:analysis}.
\section{Analysis}
\label{sec:analysis}
We perform the numerical scan in the ranges shown in Table \ref{tab:scan}.
All the parameters are introduced at the GUT scale, except for $\mu$ and $m_{A^0}$ that are defined at the EW scale.
The running (including neutrinos) is performed with the \texttt{SPheno-4.0.4} code \cite{Porod:2003um, Porod:2011nf} generated by \texttt{SARAH-4.14.3} \cite{staub2008sarah, Staub_2012, Staub_2014}.

Regarding the neutrino sector, once the neutrino Yukawas are known, a phenomenologically viable seesaw mechanism can always be realised by a suitable choice of the right-handed neutrino Majorana mass matrix \cite{Masiero:2002jn}, see Appendix \ref{appsec:Majorana}.
The running of the neutrinos is performed as detailed in \cite{Antusch:2005gp}.
As a result of the RGE, the neutrino masses and mixing can be slightly modified.
To optimize the scan, we feed the program with $Y_\nu$ at the GUT scale and the effective mass matrix for the light neutrinos at low energies as inputs.

After the running, the relevant observables are computed and the following constraints are imposed:
\begin{itemize}
    \item Neutralino LSP and radiative EWSB
    \item $m_{\chi_1^\pm} > 160\, \GeV$ \cite{ATLAS-CONF-2019-014}
    \item $m_h \in [122,\, 128]$ GeV
    \item $m_{\tilde g}\geq 2$ TeV
    \item $1.12\times 10^{-9} \leq$  BR($B_s\to \mu^+\mu^-$) $\leq 4.48\times10^{-9}$
    \item $2.79\times 10^{-4} \leq$  BR($b\to s\gamma$) $\leq 4.63\times10^{-4}$
    \item Neutrino observables within the ranges in \Tab{tab:nufit}.
    \item Tau rare decays below the current limits in \Tab{tab:LFV}.
\end{itemize}

We also check the thermal relic density of the neutralinos.
For higgsino-like LSP, the dark matter candidate is a WIMP and the predicted relic abundance is usually below the WMAP measurement \cite{Jarosik:2010iu},
\beq
    \Omega_{\rm DM}\, h^2 ~=~ 0.113\pm0.0035~ \text{ at } 68\% .
\eeq
We accept those solutions that predict a lower density and assume that additional effects due to exotic matter, for instance gravitinos, axions, moduli fields or others, can account for the rest \cite{Baer:2011ec, Baer:2012uy, Bae:2015jea, Han:2019vxi}.
\begin{table}[t!]
    \centering
    {\renewcommand{\arraystretch}{1.05}
    \resizebox{0.55\columnwidth}{!}{
    \begin{tabular}{l c c}
        \toprule
        \bf\quad MSSM sector & \bf Range & \bf Sneutrinos\qquad \\[1.5pt]
        \midrule
        \quad $m_0^{(1,2)}$ (TeV) & $[\,0,\, 15\,]$ & $m_{\tilde \nu_{1,2}}=m_0^{(1,2)}$ \qquad\\[3pt]
        \quad $m_0^{(3)}$ (TeV) & $[\,0,\, 15\,]$ & $m_{\tilde \nu_{3}}=m_0^{(3)}$ \qquad\\[3pt]
        \quad $\tan\beta$ & $[\,3,\, 60\,]$ & \\[3pt]
        \quad $m_{A^0}$ (TeV) & $[\,0,\, 10\,]$ & \\[3pt]
        \quad $\mu$ (TeV) & $[\,0,\, 1 \,]$ & \\[3pt]
        \quad $M_{1/2}$ (TeV) & $[\,0,\, 4\,]$ & \\[3pt]
        \quad $A_0$ & $[\,-4,\, 4\,]\, m_0^{(3)}$ & $A_0^\nu = A_0$ \qquad\\[3pt]
        \bottomrule
    \end{tabular}}}
    \captionsetup{width=0.8\textwidth,labelsep=none}
    \caption{\label{tab:scan} \small
    .- Parameter space of the numerical scan.}
\end{table}
\subsection{Lepton observables}
\label{subsec:CLFobs}
\begin{figure}[t!]
  \centering
    \begin{minipage}[c]{\textwidth}
        \centering
        \includegraphics[width=0.475\textwidth]{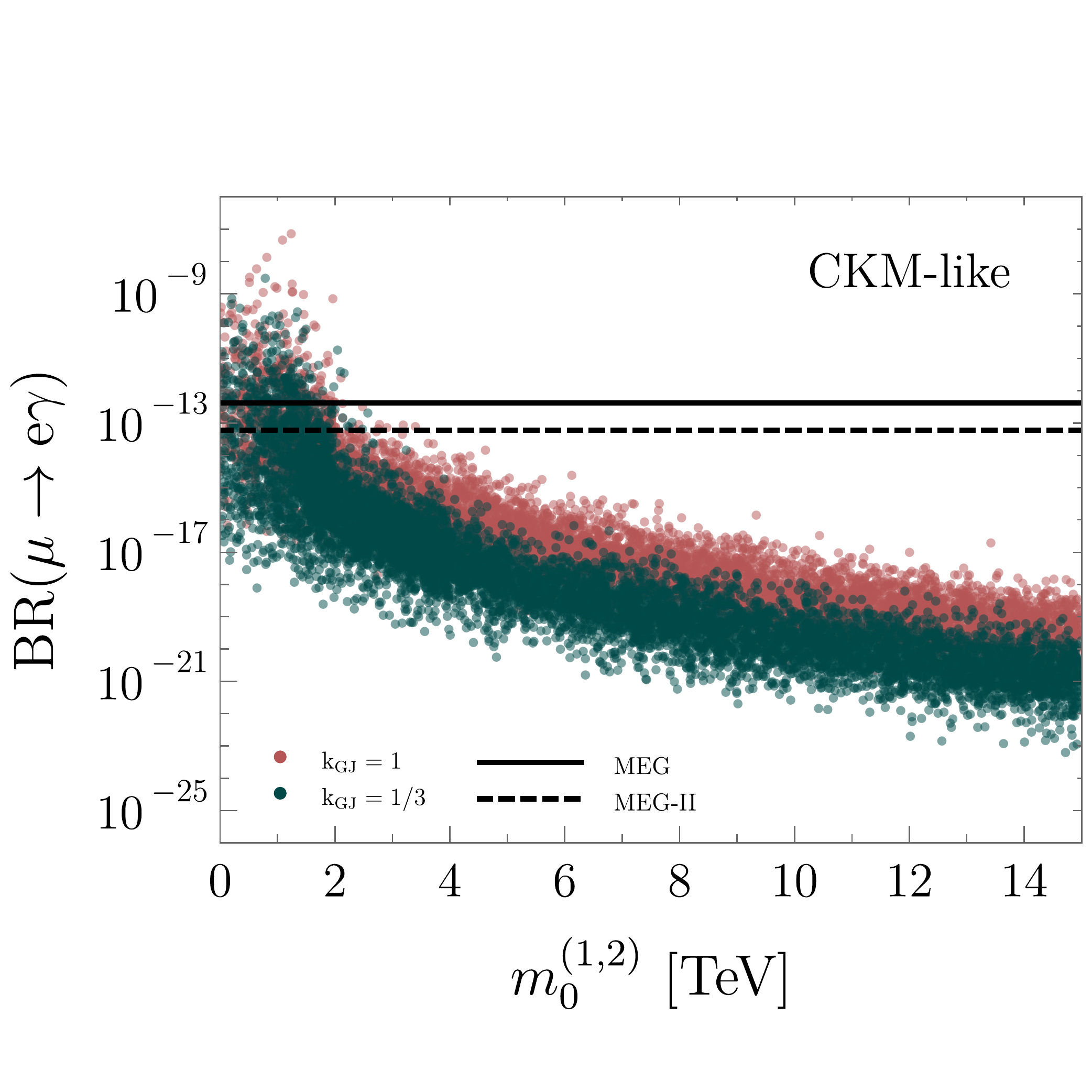}
        \hspace{0.5cm}
        \includegraphics[width=0.475\textwidth]{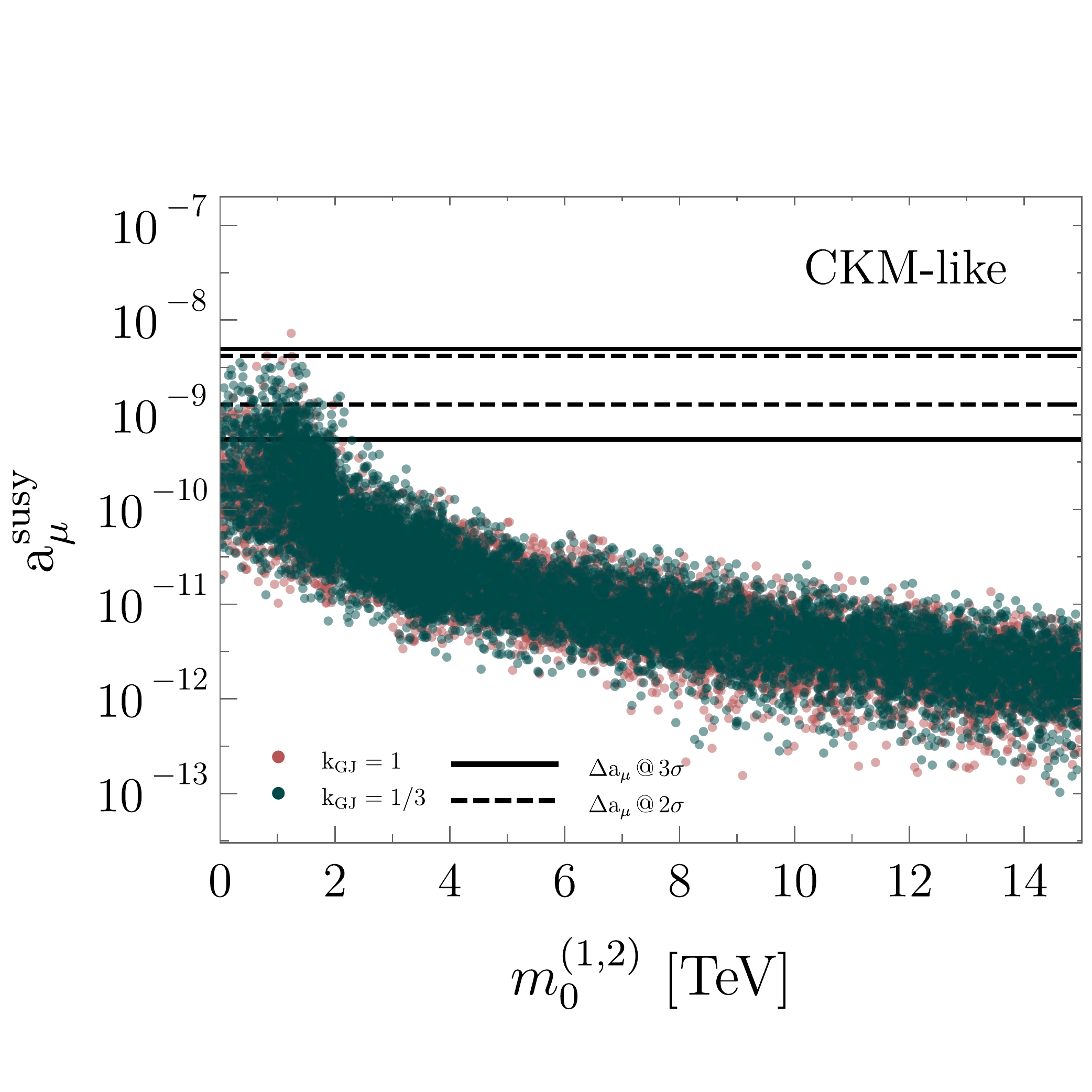}
    \end{minipage}\vspace{-1.cm}\par\vspace{\baselineskip}
    \begin{minipage}[c]{\textwidth}
        \centering
        \includegraphics[width=0.475\textwidth]{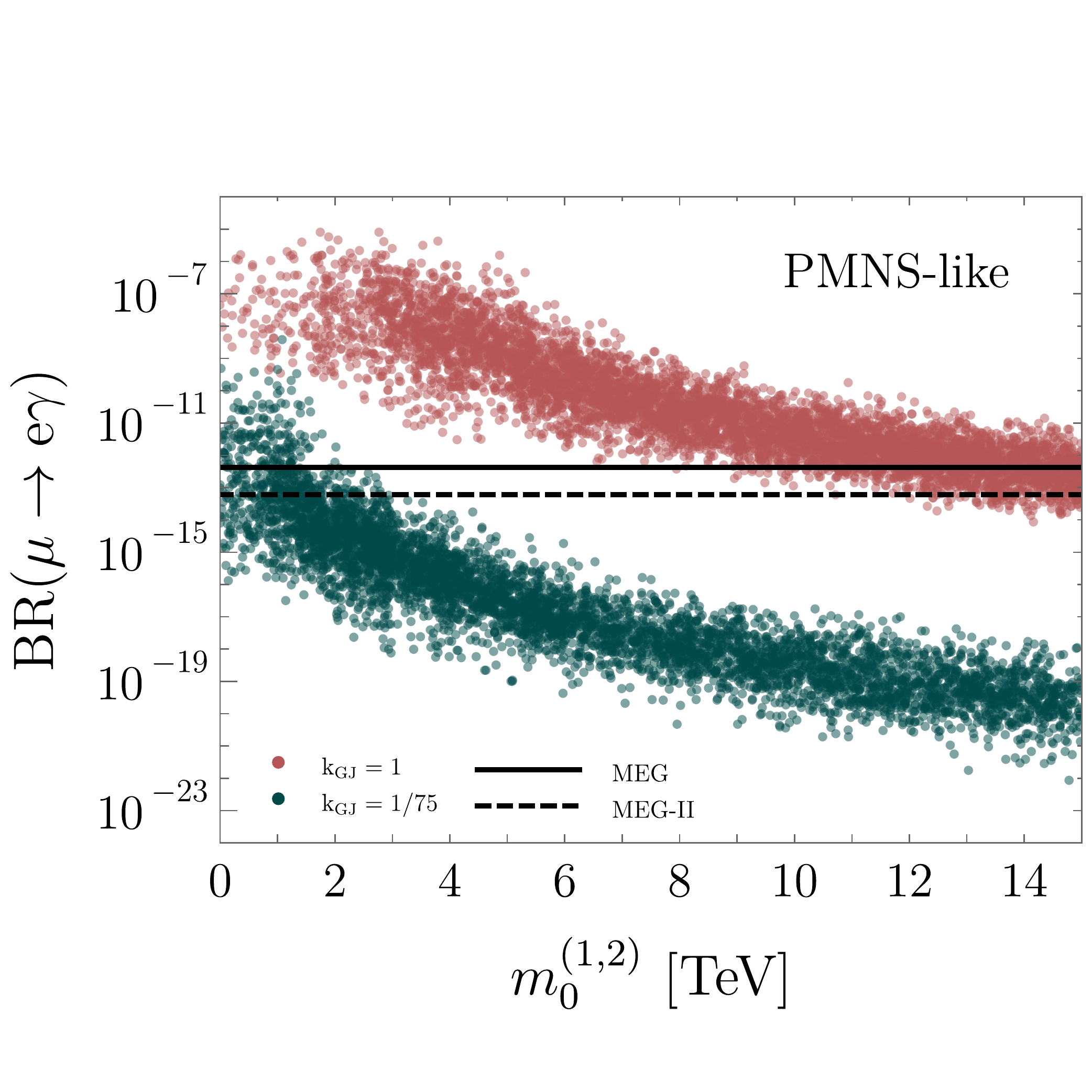}
        \hspace{0.5cm}
        \includegraphics[width=0.475\textwidth]{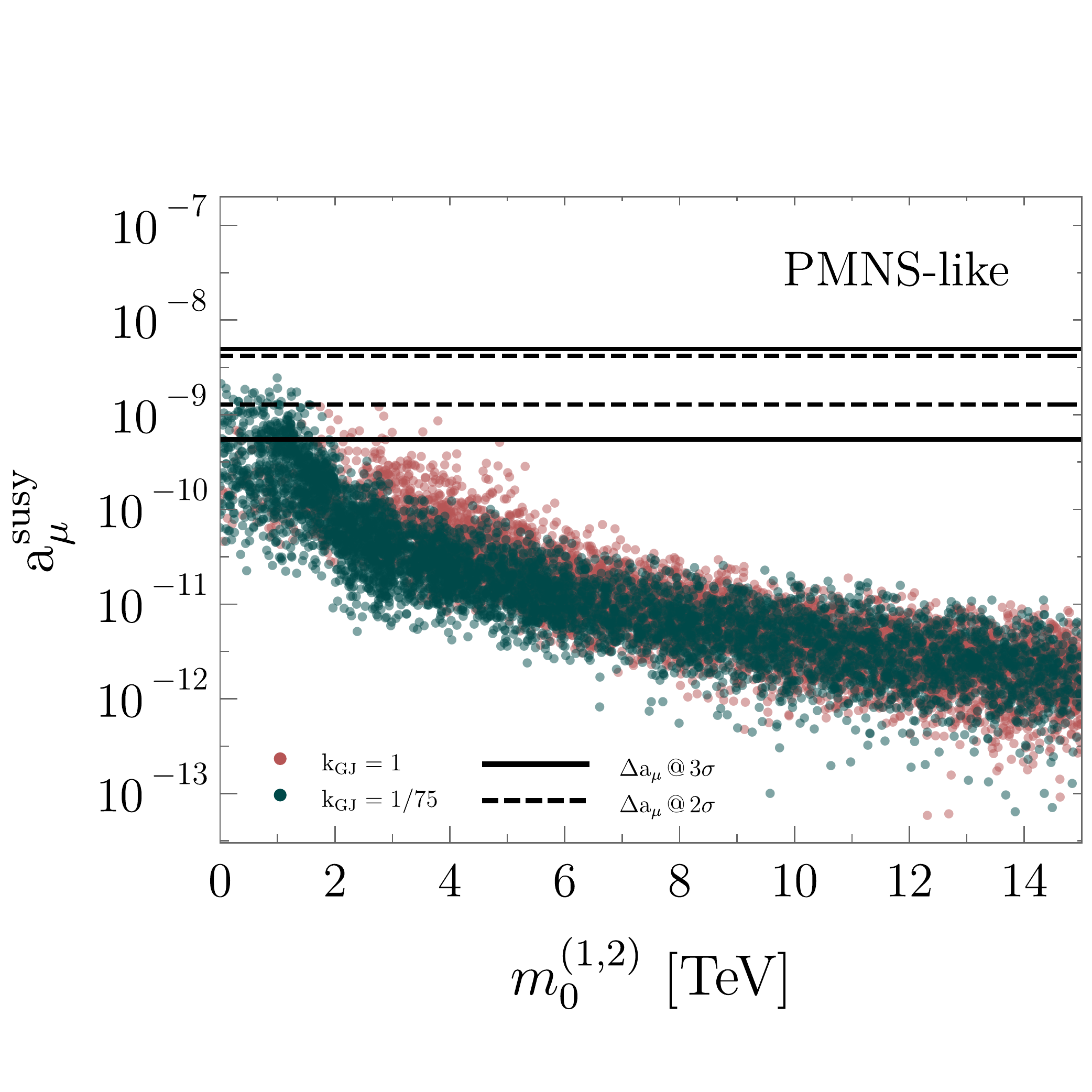}
    \end{minipage}\vspace{-0.5cm}\par\vspace{\baselineskip}
    \captionsetup{width=\textwidth,labelsep=none}
    \caption{\label{fig:versus_m012}.- \small
    {\bf Top:} ${\rm BR}(\mu\to\e\gamma)$ (left) and $\gmususy$ (right) versus the common soft mass of the first two generations of sleptons at the GUT scale in the CKM-like.
    Red/light points refer to $\kGJ=1$ while green/dark points correspond to $\kGJ\neq 1$.
    The black (dashed) lines signal the current (future) limit from \texttt{MEG (MEG-II)} \cite{TheMEG:2016wtm, Baldini:2018nnn}.
    {\bf Bottom:} The same as above but in the PMNS-like case.}
\end{figure}
\begin{figure}[t!]
  \centering
    \vspace{-0.25cm}
    \begin{minipage}[c]{\textwidth}
        \includegraphics[width=0.475\textwidth]{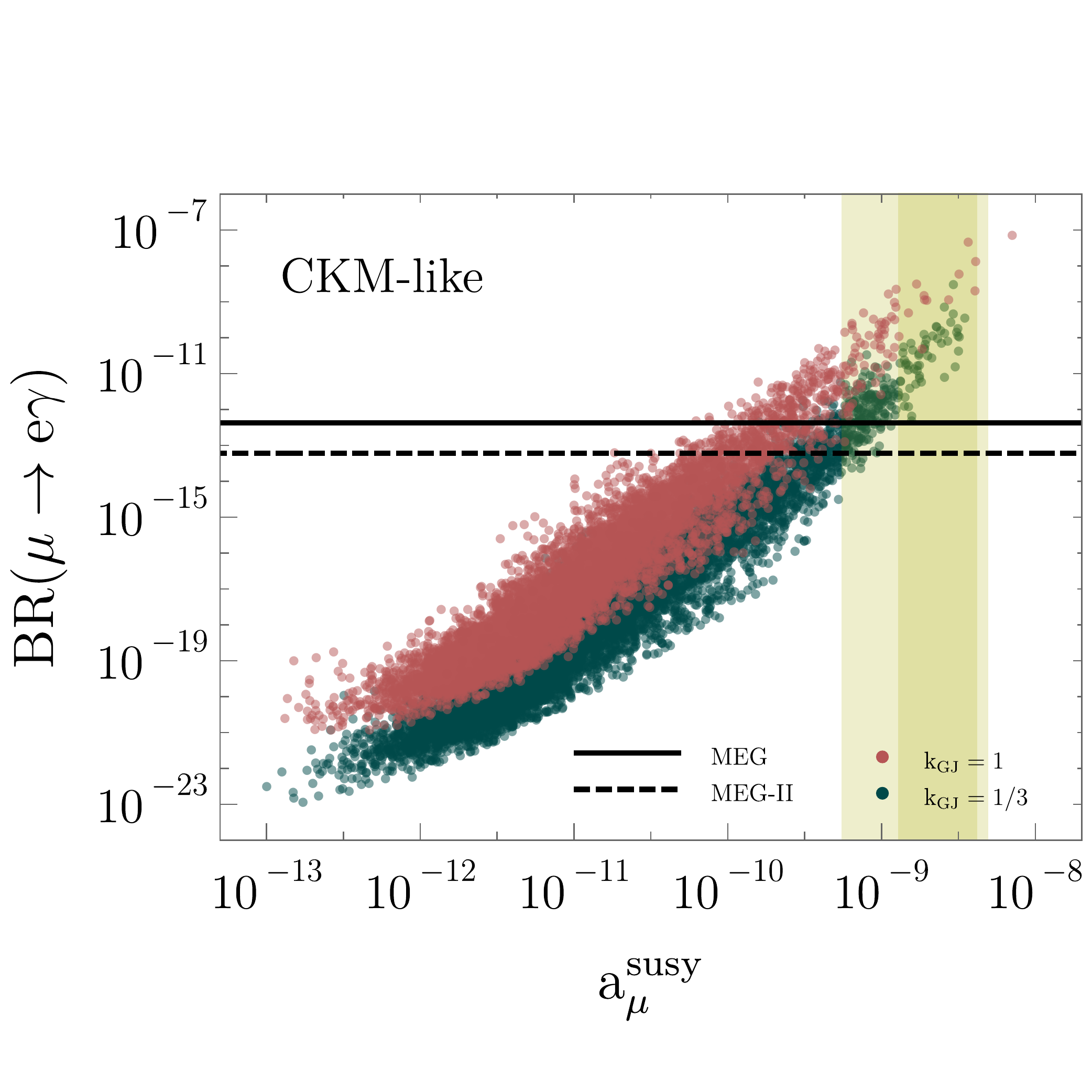}
        \hspace{0.5cm}
        \includegraphics[width=0.48\textwidth]{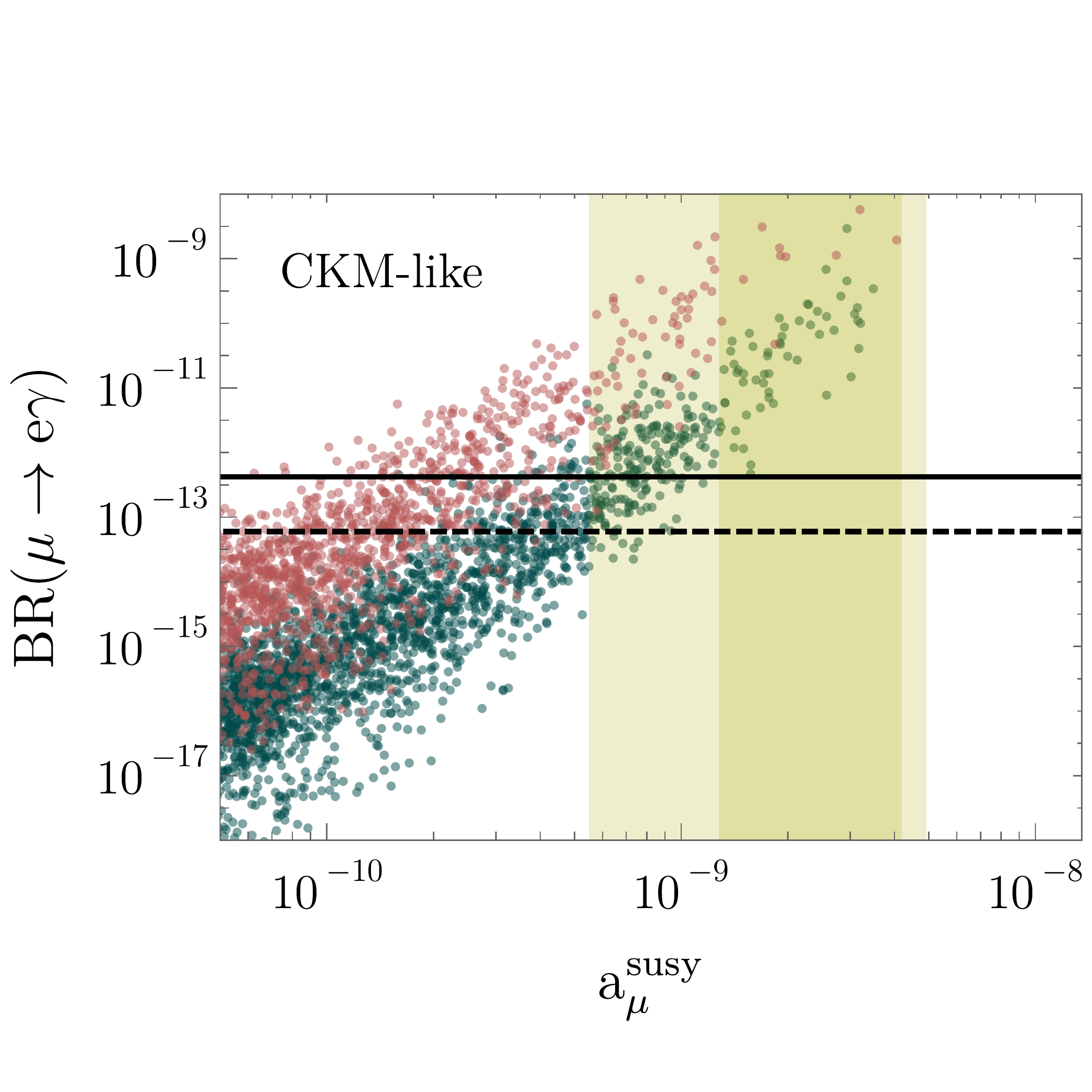}
    \end{minipage}\vspace{-1.75cm}\par\vspace{\baselineskip}
    \begin{minipage}[c]{\textwidth}
        \includegraphics[width=0.475\textwidth]{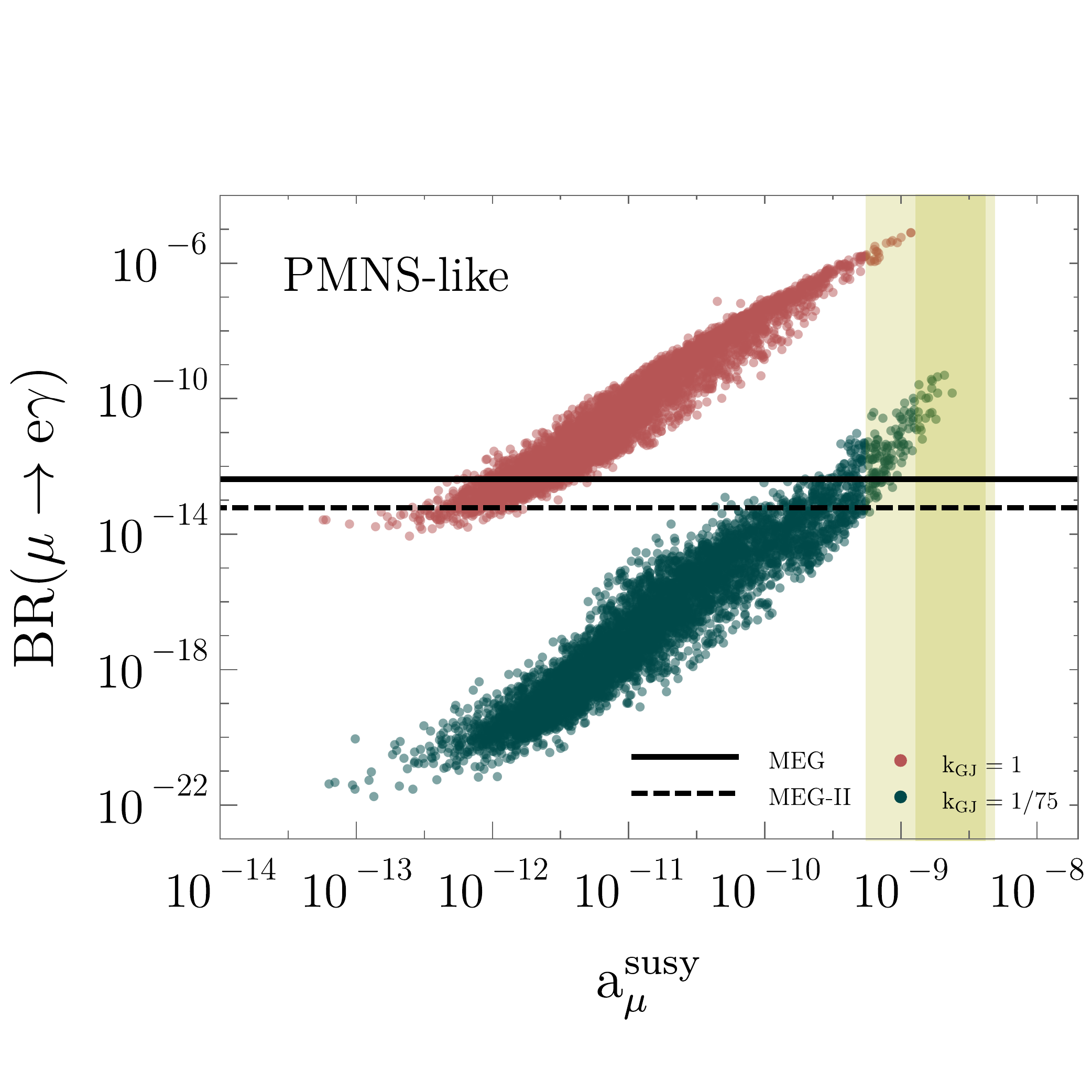}
        \hspace{0.5cm}
        \includegraphics[width=0.48\textwidth]{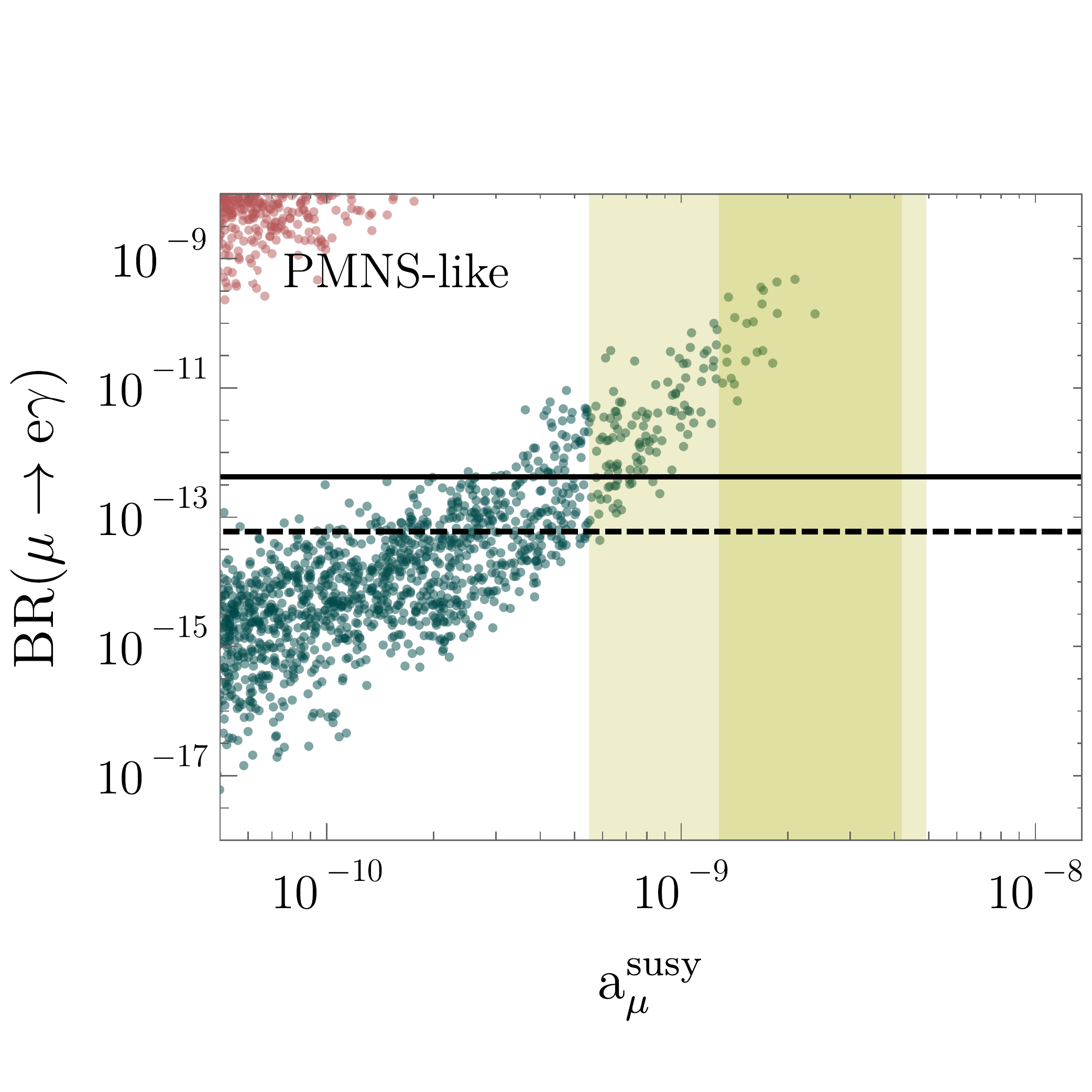}
    \end{minipage}\vspace{-1.75cm}\par\vspace{\baselineskip}
    \begin{minipage}[c]{\textwidth}
        \includegraphics[width=0.48\textwidth]{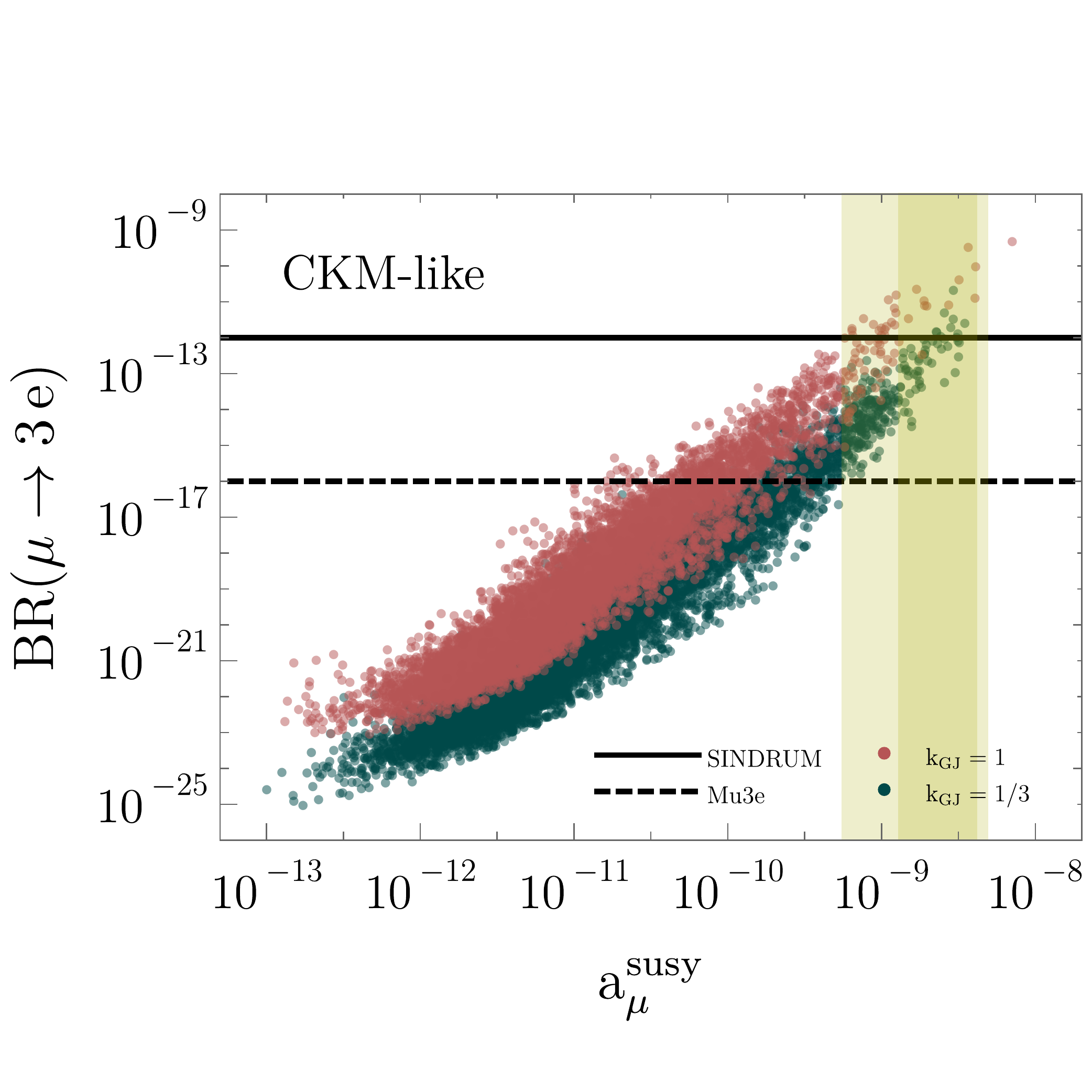}
        \hspace{0.5cm}
        \includegraphics[width=0.48\textwidth]{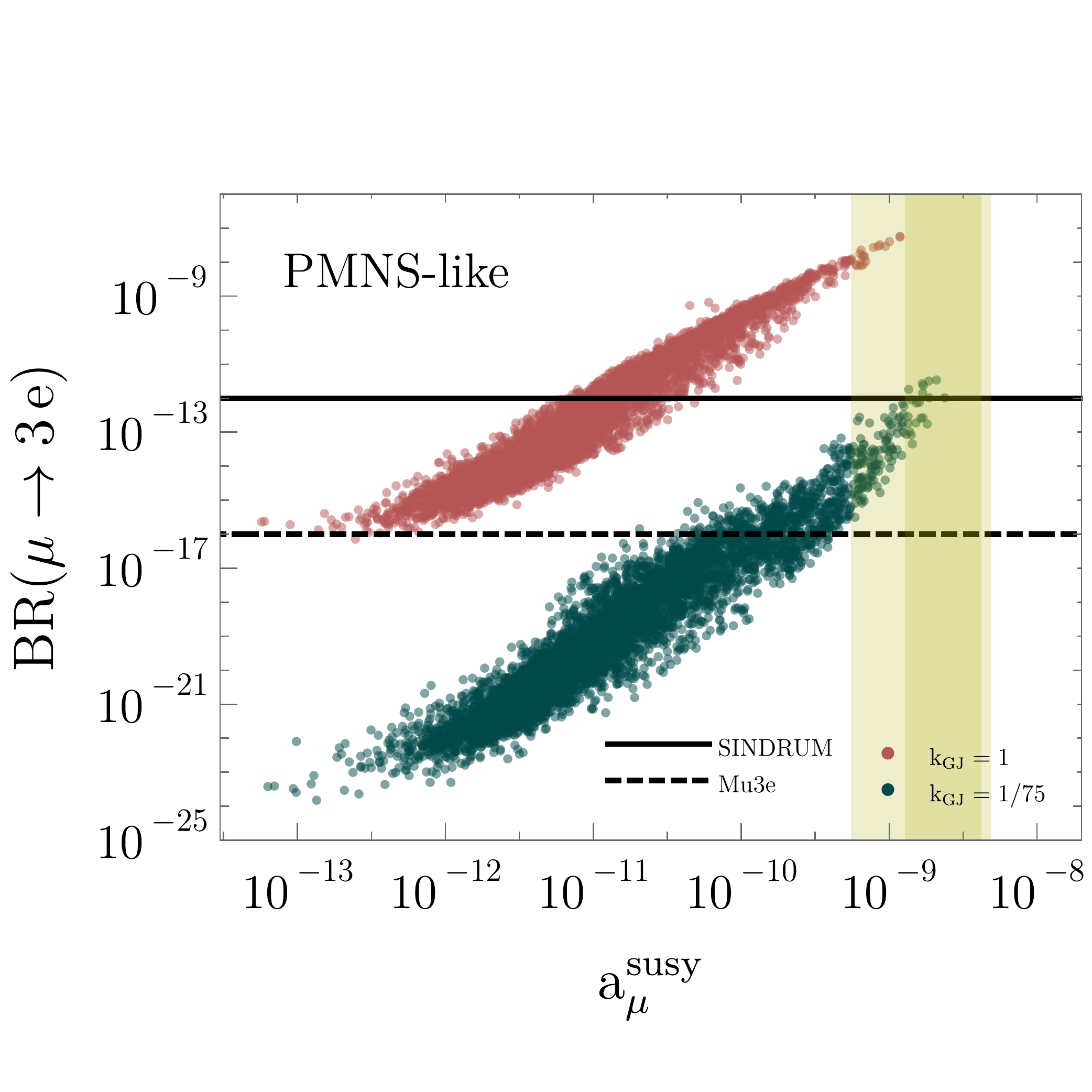}
    \end{minipage}\vspace{-1.cm}\par\vspace{\baselineskip}
    \captionsetup{width=\textwidth,labelsep=none}
    \caption{\label{fig:mey_m^0_12_g2}.- \small
    {\bf Top:} ${\rm BR}(\mu\to\e\gamma)$ versus the supersymmetric contribution to the anomalous magnetic moment of the muon in the CKM-like case.
    The right panel is a magnification of the region of interest.
    Red/light points correspond to $kGJ=1$ while green/dark points to $\kGJ\neq 1$.
    The black (dashed) line signals the current (future) limit from \texttt{MEG (MEG-II)}.
    {\bf Middle:} The same as before but in the PMNS-like case.
    {\bf Bottom:} ${\rm BR}(\mu\to3\e)$ versus the supersymmetric contribution to the anomalous magnetic moment of the muon in the CKM-like (left) and PMNS-like (right) cases.
    Red/light points correspond to $kGJ=1$ while green/dark points to $\kGJ\neq 1$.
    The black (dashed) line signals the current (future) limit from \texttt{SINDRUM (Mu3e)}
    }
\end{figure}

First, we analyse the dependency of the two transitions, $\mu\to\e$ and $(g-2)_\mu$, on the mass of the sparticle that enter the loop.
As discussed in Section \ref{subsec:g2} and \ref{subsec:CLFV}, both amplitudes are inversely proportional to the muonic sneutrino mass, see Eqs.\eqref{eq:g2MIA} and \eqref{eq:mey_AL_MIA}.
We corroborate this result in \Fig{fig:versus_m012}, where the values of ${\rm BR}(\mu\to\e\gamma)$ and $\gmususy$ have been plotted versus $m^{(1,2)}_0$, the parameter that controls the mass of the first two generations of sfermions.
The first (second) row corresponds to the CKM-like (PMNS-like) case.
The left column is for ${\rm BR}(\mu\to\e\gamma)$ and the right column for $\gmususy$.

As expected, the largest branching fraction and magnetic moment are observed for $m_0^{(1,2)}\lesssim 2$ TeV.
The CKM-like case exhibits, in general, smaller values than the PMNS-like models.
This is totally reasonable since the off-diagonal elements of the former are produced by the CKM matrix (almost diagonal) whilst the seconds are dictated by the PMNS matrix with larger mixing angles, see \Eqs{eq:ckmlike_a}{eq:pmnslike_a}.
The colours red/light and green/dark refer to $\kGJ=1$ and $\kGJ\neq 1$, respectively.
We observe that the amplitude for the FC process decrease when the $\kGJ$ factor is reduced while, the same variation in $\kGJ$, has little effect on $\gmususy$.
This is in agreement with Eqs.~\eqref{eq:g2MIA}, \eqref{eq:mey_AL_MIA} and \eqref{eqn:RGEsoftmass}.

In \Fig{fig:mey_m^0_12_g2}, the results for ${\rm BR}(\mu\to\e\gamma)$ and $a_\mu^{\rm susy}$ are compared.
Again, the colour legend is associated with $\kGJ=1$ (red/light) and $\kGJ\neq 1$ (green/dark).
The first row shows the distribution of points for the CKM-like case.
In the left column, the full range of points is represented whereas, in the right column, a magnification of the region of interest can be found.
The second row is for the PMNS-like case and shows similar plots.
In the case where $\kGJ=1$, both scenarios are ruled out by the \texttt{MEG} limit, since they predict a larger branching fraction for the process $\mu\to\e\gamma$.
As discussed in Section \ref{subsec:g2vsmey}, a $\kGJ$ factor that introduces a splitting between the up-type quarks and neutrinos may reduce the discrepancy and make possible to conciliate all the experimental limits.

For the CKM-like case, the value $\kGJ=1/3$ is enough to fulfill the current bound from \texttt{MEG} and the expected limit from \texttt{MEG-II}.
This is in agreement with what was observed in \Fig{fig:mey_m^0_12_g2}.
In contrast, the PMNS-like case requires a significantly smaller value, $\kGJ\sim 1/75$.
This is consistent with Eq.~\eqref{eqn:RGEsoftmass}, which indicates that the generated off-diagonal elements should be larger in the PMNS case, although we obtain a slightly smaller $\kGJ$ than our estimate in section~\ref{subsec:g2vsmey}.
These tiny values of $\kGJ$ would allow a supersymmetric explanation of the discrepancy with PMNS-like mixings, but they are not natural in a GUT scenario. In Eq.~\eqref{eq:pmnslike_a}, we are assuming that the representation with a Georgi-Jarlskog factor gives the dominant contribution to the neutrino Yukawa couplings.
However, a $\kGJ\sim 1/75$ would require all other contributions to be absent or more than two orders of magnitude smaller.

On the other hand, we would like to emphasize that a smaller discrepancy of the experimental results from the SM predictions would enlarge the portion of the parameter space where both observables are consistent.
In particular, some points in the CKM-like case with $\kGJ=1$ may be permitted.
This is the case if the discrepancy on the anomalous magnetic moment is reduced due to a slight decrease of the experimental central value \cite{Grange:2015fou} or because of an improvement in the SM calculation in the direction of the measured quantity \cite{Borsanyi:2020mff}.

The last row of Figure \ref{fig:mey_m^0_12_g2} is dedicated to the CLFV decay $\mu\to3\e$.
Although the present limit for the process (black, continuous line) is less restrictive that the correspondent to $\mu\to\e\gamma$, the expected sensitivity (black, dashed line) will become more limiting.
Specifically, the region of points where $\kGJ\neq 1$ and $\gmususy$ is compatible with the experimental measurement at $3\sigma$ or more, will be totally scrutinised.
In other words, if the discrepancy in the anomalous magnetic moment of the muon is due to supersymmetric corrections like the ones analysed here, a signal in the $\mu\to 3\e$ channel is expected within the $\sim 10^{-12}-10^{-16}$ range.
Similar conclusions are extracted from the inspection of the observable ${\rm CR}(\mu{\rm N}\to\e{\rm N})$.

\subsection{Benchmark point}
\label{subsec:spectrum}
\begin{figure}[t!]
  \centering
    \vspace{-0.75cm}
    \begin{minipage}[c]{\textwidth}
        \centering
        \includegraphics[width=0.6\textwidth]{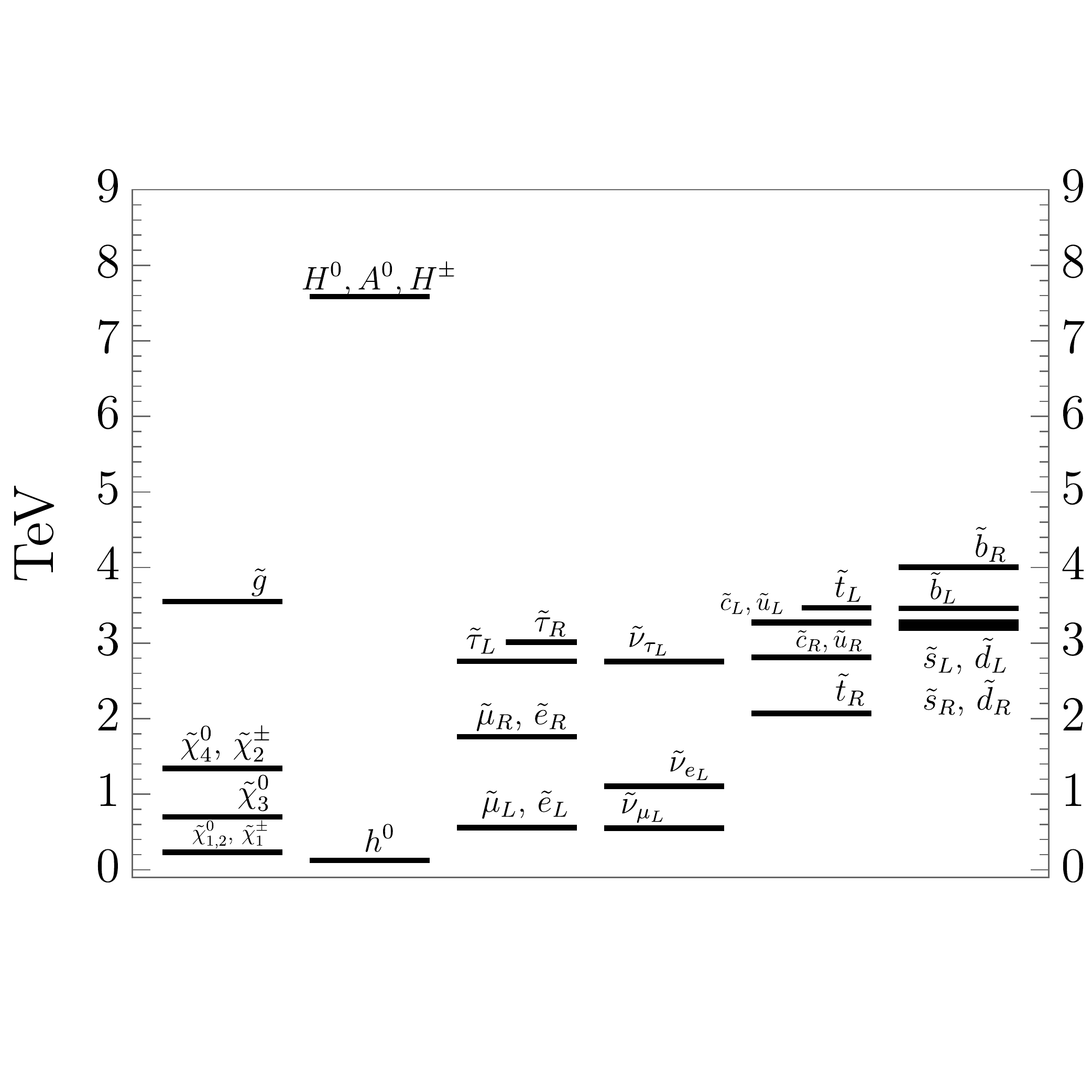}
    \end{minipage}\vspace{-2.cm}\par\vspace{\baselineskip}
    \captionsetup{width=\textwidth,labelsep=none}
    \caption{\label{fig:spect}.- \small 
    Spectrum for the benchmark point in \Tab{tab:BenchPoint}.}
\end{figure}
\begin{table}[h!]
    \centering
    {\renewcommand{\arraystretch}{1.5}
    \resizebox{0.85\columnwidth}{!}{
    \begin{tabular}{c c c c c c c}
        \toprule
        $\bm{m_0^{(1,2)}}~ [\TeV]$ & $\bm{m_0^{(3)}}~ [\TeV]$ & $\bm{M_{1/2}}~ [\TeV]$ & $\bm{A_0}~ [\TeV]$ & $\bm{\mu}~ [\TeV]$ & $\bm{m_{A^0}}~ [\TeV]$ & \boldmath $\tan\beta$ \\[2pt]
        \midrule
        $0.617$ & $3.034$ & $1.636$ & -$4.102$ & $0.220$ & $7.584$ & $21.7$ \\
        \bottomrule
    \end{tabular}}}
    \captionsetup{width=0.8\textwidth,labelsep=none}
    \caption{\label{tab:BenchPoint}.- \small
        Input parameters for the benchmark point.}
\end{table}
In this section we discuss the main phenomenological features of a representative benchmark point that satisfies all the constraints detailed before, reproduces $\Delta a_\mu$ at $2.5\sigma$ and is in agreement with \texttt{MEG} limits.
The input parameters are presented in \Tab{tab:BenchPoint}, which belong to the CKM-like case.
The produced spectrum is depicted in \Fig{fig:spect}.

As commented throughout the text, the models analysed here with $\mu\ll M_{1/2}$ predict lightest neutralinos and chargino which are higgsino-like and, therefore, almost degenerate in mass.
Because of that, the two-body decay of $\tilde \chi_2^0$ and $\tilde \chi_1^\pm$ to $\tilde \chi^0_1$ is forbidden and LHC limits for this compressed spectrum are rather weak.
The third neutralino is mostly bino whilst the forth neutralino and second chargino are wino-like.
As expected from models with universal gaugino masses at the GUT scale, the gluino is the heaviest superpartner in the gauge sector and, in our case, a mass of $\sim 3.5\, \TeV$ is predicted.

The sfermion sector displays the following hierarchy among generations: the lightest states for both sleptons and squarks belong to the first two generations whereas the third generation is associated with the heaviest states.
This is consistent with $m_0^{(1,2)}\ll m_0^{(3)}$.
This effect is much stronger in sleptons than in squarks, where the masses are dominated by the gluino mass and the difference in $m_0^{(1,2)}\ll m_0^{(3)}$ is less relevant.
The only exception to this pattern is the lightest stop (a right-handed stop) that turns out to be the lightest squark.

The main effect of the RGE on the sfermion masses is due to the gauge interactions, which increase the mass as they are evolved down to the EW scale, and, mostly for the third generation, the Yukawa couplings and trilinears, which decrease their masses.
Thus, in general, heavier left-handed states are expected for the first two generations.
For up-type squarks, this is also the case in the third generation since the right-handed stop contribution that lowers its mass is proportional to $2y_t^2$ while the term associated with the left-handed superpartner goes as $y_t^2+y_b^2$.
These two effects make the right-handed stop the lightest squark.
For down-type squarks, the situation is reversed since the left-handed sbottom receives corrections proportional to $y_t^2+y_b^2$ while the right-handed sbottom contribution is $\propto 2y_b^2$.
This effect is, therefore, opposite to the gauge contribution and, in this particular case, makes the left-handed sbottom lighter than the right-handed one.
\begin{figure}[t!]
  \centering
    \vspace{-0.25cm}
    \begin{minipage}[c]{\textwidth}
        \includegraphics[width=0.475\textwidth]{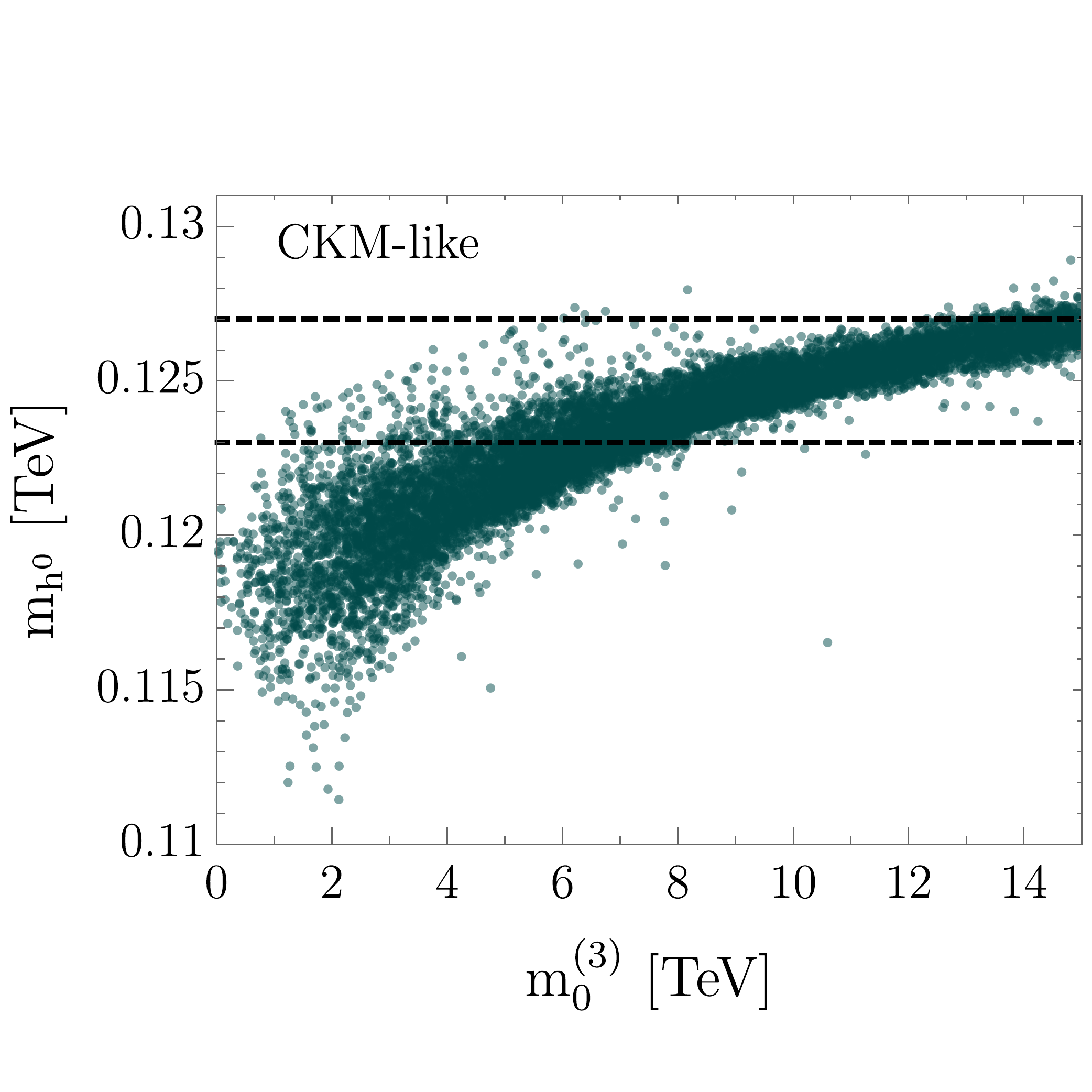}
        \hspace{0.5cm}
        \includegraphics[width=0.475\textwidth]{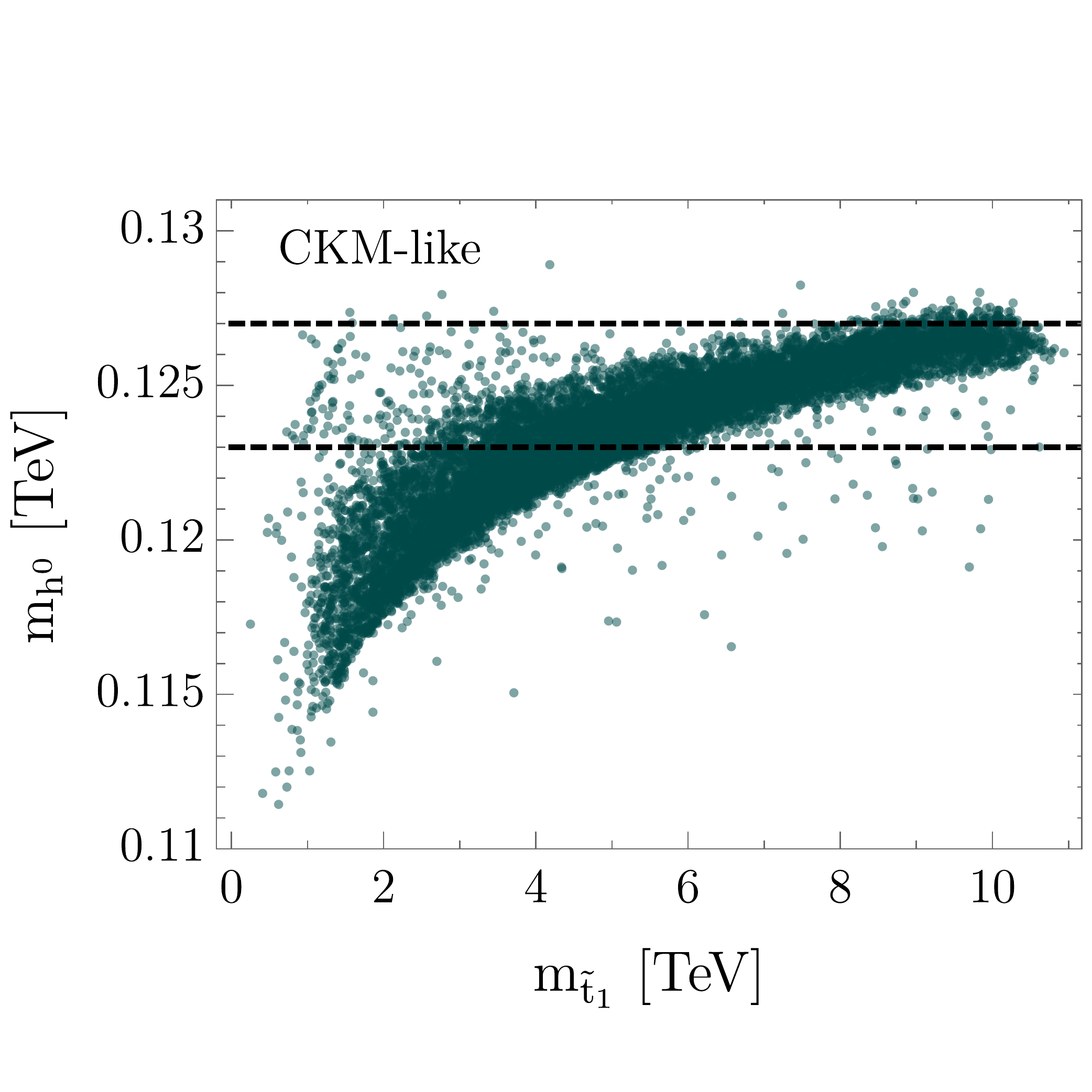}
    \end{minipage}\vspace{-1.25cm}\par\vspace{\baselineskip}
    \captionsetup{width=\textwidth,labelsep=none}
    \caption{\label{fig:Higgs}.- \small 
    Higgs mass versus the input parameter at the GUT scale $m_0^{(3)}$ (left) and the mass of lightest stop at the SUSY scale (right).}
\end{figure}

The situation is slightly more involved in the slepton sector, where the combined effect of $m_0^{(1,2)}\ll m_0^{(3)}$ and up-like neutrino Yukawas interfere in the evolution of the left-handed sleptons and make them lighter than the right-handed superpartners, also for the first two generations.
This hierarchy is indeed congruent with the explanation of $\Delta a_\mu$, since they are the ones mediating the transitions $\mu \to \e \gamma$ and $\mu \to \mu \gamma$.
Left-handed sneutrinos follow the same pattern than their charged counterpart and satisfy $m_{\tilde \nu_\ell} \simeq m_{\tilde \ell}$.

We check that the predicted spectrum is compatible with LHC direct searches.

Regarding the Higgs sector, we obtain a decoupled spectrum where the lightest Higgs is SM-like and the other CP-even, CP-odd and charged states are much heavier and quasi-degenerate in mass.
We would like to emphasize that it is possible to reproduce the observed Higgs boson mass and $\Delta a_\mu$ simultaneously in these realisations.
In \Fig{fig:Higgs}, we plot the mass of the lightest Higgs boson against $m_0^{(3)}$ (left) and the lightest stop mass (right) in the CKM-like case\footnote{Similar results are found for the PMNS-like case.}.
Although marginal points that reproduce the Higgs boson mass can be found for lower values of $m_0^{(3)}$ and $m_{\tilde t_1}$, the bulk of the scan indicates that
\beq
    m_0^{(3)} \gtrsim 4\, \TeV,\quad m_{\tilde t_1},\, m_{\tilde t_2} \gtrsim 2.5\, \TeV
\eeq
are preferred to produce a lightest Higgs state near the experimental mass.
In conclusion, $\Delta a_\mu$ and the Higgs mass are accomplished simultaneously for non-universal models where distinct scalar soft masses are assumed for the first-two and third generation.
The lightness of the former allows for a suitable supersymmetric contribution that accounts for the measured discrepancy in the muon anomalous magnetic moment whereas stops within the few-TeV regime are enough to produce a Higgs mass at the observed central value, $125\, \GeV$.

\section{Conclusions}
\label{sec:conclusion}
We have investigated how to explain $\Delta a_\mu$ in supersymmetric models with right-handed neutrinos while keeping indirect flavour bounds under control.

In Section \ref{sec:seesaw}, the main ingredients of a supersymmetric type-I seesaw model were presented together with some relations that arise between the Yukawa couplings of different species of sfermions within GUT realisations.
In particular, we discussed how a Georgi-Jarlskog factor between the Yukawa couplings of up-type quarks and neutrinos may make the latter significantly smaller than the former.
We provide an specific example where this happens.
As exposed in this section, we focus our analysis on GUT realisations where non-universal conditions for the scalar soft masses are assumed at the unification scale.
This kind of models are favoured from naturalness considerations since they succeed at reproducing the observed Higgs mass value while producing a low amount of fine-tuning, providing that the Higgs mass parameter is $\mu\lesssim 1\, \TeV$.
A direct consequence of this condition is that the lightest neutralinos and charginos are higgsino-like and therefore quasi-degenerate in mass.
The resulting phenomenology is poorly constrained from direct searches at LHC, so that it remains a highly attractive scenario to explore in the next generation of linear colliders.

In section \ref{sec:RGE}, we recalled a well-known fact: right-handed neutrinos in SUSY models induce off-diagonal elements in the slepton soft mass matrices during the running down to the EW scale through the RGEs.
The impact of the these terms in the flavour-changing transitions $\ell_i\to\ell_j$ has been investigated in Section \ref{sec:obs}.
Analytical expressions have been derived together with a simple relation between the amplitudes of $\gmususy$ and ${\rm BR}(\mu\to\e\gamma)$ that holds for our models.

Section \ref{sec:analysis} was dedicated to our numerical scan.
We observe that non-universal models allow for a Higgs boson mass at $125\, \GeV$ and a suitable contribution to the anomalous magnetic moment of the muon which can account for the discrepancy between the experimental and the theoretical SM value.
For those solutions, charged-lepton flavour-violating processes remain below the experimental limits if the neutrino Yukawas at the GUT scale are suppressed compared to the up-type quark couplings.
Due to our little knowledge about the neutrino couplings, this is possible even within a GUT unification context, with a Georgi-Jarlskog factor, $\kGJ$, as explained in Section \ref{sec:seesaw}.
The required factor between up and neutrino Yukawas depends on the mixing associated with the left-handed neutrinos, small (CKM-like) or large (PMNS-like).
While, in the first case a $1/3$ factor, which can be easily accomodated within a complete theory of flavour, is sufficient, the PMNS-case requires a very small factor, which might be perceived as quite unnatural.
Nonetheless, a future change in the experimental results could be favourable for these unified models, if it reduces the discrepancy with the SM predictions.
Specifically, in this case, a strict unification between the Yukawa couplings of up-type quarks and neutrinos could be reached if the left-handed neutrino Yukawas have a CKM-like mixing.

\section{Acknowledgements}
\label{sec:acknow}
AM acknowledges support from La-Caixa-Severo Ochoa scholarship.
AM and OV are supported by Spanish and European funds under MICIU Grant FPA2017-84543-P and by the “Centro de Excelencia Severo Ochoa” programme under grant SEV-2014-0398. OV acknowledges partial support from the “Generalitat Valenciana” grant PROMETEO2017-033.
JMY acknowledges funding from the National Natural Science Foundation of China (NNSFC) under grant Nos.11675242, 11821505, and 11851303, from Peng-Huan-Wu Theoretical Physics Innovation Center (11947302), from the CAS Center for Excellence in Particle Physics (CCEPP), from the CAS Key Research Program of Frontier Sciences and from a Key R\&D Program of Ministry of Science and Technology under number 2017YFA0402204.
LW acknowledges funding from the National Natural Science Foundation of China (NNSFC) under grant No. 11705093.

\appendix

\section{Loop Functions}
\label{appsec:loop}
\bea
    f_1^{(c)}(x) & = & \frac{2 \,+\, 3x \,-\, 6x^2 \,+\, x^3 \,+\, 6x\log{x}}{6(1-x)^4}
    \label{appeq:fc1} \\
    f_2^{(c)}(x) & = & \frac{-3 \,+\, 4x \,-\, x^2 \,-\, 2\, \log{x}}{(1-x)^3} 
    \label{appeq:fc2} \\
    f_3^{(c)}(x) & = & \frac{-5 \,+\, 4x \,+\, x^2 \,-\, 2\, (1+2x)\log{x}}{(1-x)^4} 
    \label{appeq:fc3} \\
    f_1^{(n)}(x) & = & \frac{1 \,-\, 6x \,+\, 3x^2 \,+\, 2x^3 \,-\, 6x^2\log{x}}{6(1-x)^4} 
    \label{appeq:fn1} \\
    f_2^{(n)}(x) & = & \frac{1 \,-\, x^2 \,+\, 2x\log{x}}{(1-x)^3} \label{appeq:fn2} \\
    f_3^{(n)}(x) & = & \frac{1 \,+\, 4x \,-\, 5x^2 \,+\, 2x(2+x)\log{x}}{(1-x)^4} \label{appeq:fn2}
\eea

\section{\boldmath MIA expressions for $\gmu$ and $\mu\to\e\gamma$}
\label{appsec:MIA_g2_m2y}
We follow the conventions in \cite{Hisano:1995cp} for the neutralino-sfermion-fermion and chargino-sfermion-fermion couplings:
\bea
    N^L_{i,\, al} & = & -\frac{g_2}{\sqrt{2}}\left[\, 
        \frac{m_{\ell_i}}{m_W\cos\beta}(O_N)_{a3}U^\ell_{li} \,+\, 
        2(O_N)_{a1}\tan\theta_W U^\ell_{l\,i+3} \,\right], \label{appeq:NL} \\[5pt]
    N^R_{i,\, al} & = & \frac{g_2}{\sqrt{2}}\, \left[\, \bigg((O_N)_{a2} + 
        (O_N)_{a1}\tan\theta_W\bigg) U^\ell_{li} \,-\,  \frac{m_{\ell_i}}{m_W\cos\beta}
        (O_N)_{a3}U^\ell_{l\,i+3}\,\right], \label{appeq:NR}\\[5pt]
    C^L_{i,\, al} & = & \frac{g_2\, m_{\ell_i}}{\sqrt{2}\, m_W\, \cos\beta}\, 
        \left(O_L\right)_{a2}\, U^\nu_{li}, \label{appeq:CL} \\[5pt]
    C^R_{i,\, al} & = & -g_2\, \left(O_R\right)_{a1}\, U^\nu_{li},
        \label{appeq:CR}
\eea
where $O_N$ is the unitary matrix that diagonalises the neutralino mass matrix, $O_L$ and $O_R$ are the unitary matrices that diagonalise the chargino mass matrix (which, in general, is not hermitian but complex) through a biunitary transformation and $U^{\ell,\, \nu}$ are the mixing matrices for charged sleptons and sneutrinos.

\subsection{Flavour-expansion theorem}
\label{appsubsec:MIA}
The flavour-expansion theorem \cite{Dedes:2015twa}, on which the mass insertion approximation is based \cite{Hall:1985dx, Gabbiani:1996hi, Buras:1997ij, Misiak:1997ei}, states that for any hermitian matrix that is diagonalized by an unitary matrix as
\beq
    U\, M^2\, U^\dagger ~=~ D ~\equiv~ \text{Diag}(m_1^2,\dots,m_n^2),
\eeq
then any real analytic function of D, which must be analytic around zero and should be possible to express in terms of its McLaurin series, can be expanded as
\beq \label{appeq:FET}
    \left[ U^\dagger\, f(D)\, U\right]_{AB} ~=~ \delta_{AB}\, f(M^2_{AA}) \;+\; M^2_{AB}\, \frac{f\left( M^2_{AA}\right)\,-\, f\left(M^2_{BB}\right)}{M_{AA}^2 \,-\, M_{BB}^2} \;+\; \cdots.
\eeq
For small enough off-diagonal elements in $M^2$, the sum in the right-hand side can be fairly good approximated by the first terms.
From \Eq{appeq:FET}, one may also derive the following relation for fermionic amplitudes \cite{Dedes:2015twa}:
\beq \label{appeq:FET_fermions}
    \sum_i\, V_{Bi}\, m_i\, f(m_i^2)\, U^*_{Ai} ~=~ \sum_C\, M_{BC}\, \left[f(M^\dagger M)\right]_{CA} ~=~ \sum_C\, \left[f(M M^\dagger)\right]_{BC}\, M_{CA},
\eeq
where $M$ is the mass matrix that is diagonalized by two unitary matrices as
\beq
    V^\dagger\, M\, U ~=~ D ~=~ {\rm Diag}(m_1,\dots,m_n).
\eeq
and
\beq
    \left[f(X)\right]_{AB} ~=~ \frac{f\left( X_{AA}\right)\,-\, f\left(X_{BB}\right)}{X_{AA} \,-\, X_{BB}}
\eeq
With this prescription, a suitable expression for $\gmususy$ and ${\rm BR}(\ell_j \to \ell_i \gamma)$ can be computed.

\subsection{\boldmath ${\rm BR}(\mu\to\e\gamma)$ and $(g-2)_\mu^{\rm susy}$}
\label{appsubsec:mey_g2_MIA}
By Lorentz invariance, the dipole amplitude for the transition $\ell_j\to\ell_i\gamma$ can be written as
\beq \label{appeq:dipole_op}
    {\cal M}_{ij} ~=~ e\, m_{\ell_j}\, \vep_\alpha(q)\, \bar u_{\ell_i}(p-q) \;\left[\, i \sigma_{\alpha\beta}\, q^\beta \left(a_L^{ji}\, P_L \;+\; a_R^{ji}\, P_R\right) \,\right]\; u_{\ell_j}(p).
\eeq
The decay rate of the process $\mu\to\e\gamma$ is then given by
\beq \label{appeq:Gamma_mey}
    \Gamma\left(\mu \to \e \gamma \right) ~=~ \frac{e^2}{16\pi}\, m_{\ell_j}^5 \left(\, \left|a_L^{\mu\e}\right|^2 \,+\, \left|a_R^{\mu\e}\right|^2 \,\right),
\eeq
while the anomalous magnetic moment of the muon is described by
\beq \label{appeq:gmu}
   a_\mu ~\equiv~ \frac{(g-2)_\mu}{2} ~=~ m_\mu^2 \left(\, a_L^\mu \,+\, a_R^{\mu}\, \right).
\eeq
In the MSSM, the amplitudes in \Eqs{appeq:Gamma_mey}{appeq:gmu} contain two terms, one for the sneutrino-chargino loop and another for the charged slepton-neutralino loop
\bea \label{appeq:a_LR_ji}
    a_{L, R}^{ji} ~=~ a_{L, R}^{(n)\, ji} \;+\; a_{L, R}^{(c)\, ji}.
\eea
When $i\neq j$, the amplitudes in \Eq{appeq:Gamma_mey} are
\bea
    a_L^{(n)\, ji} & = & \frac{1}{32\pi^2} \sum_{l,a}\, \frac{1}{m_{\tilde \ell_l}^2} \bigg[\,
        N^L_{i,\, al}\, N^{L^*}_{j,\, al}\; f^{(n)}_1(x_{al}) \:+\: 
        \frac{m_{\chi^0_a}}{m_{\ell_j}}\, N^L_{i,\, al}\, N^{R^*}_{j,\, al}\; 
        f^{(n)}_2(x_{al})\,\bigg], \label{appeq:aLn_mey} \\
    a_L^{(c)\, ji} & = & -\frac{1}{32\pi^2} \sum_{l,\,a}\, \frac{1}{m_{\tilde \nu_l}^2} 
        \bigg[\, C^L_{i,\, al}\, C^{L^*}_{j,\, al}\; f^{(c)}_1(x_{al}) \:+\: 
        \frac{m_{\chi^-_a}}{m_{\ell_j}}\, C^L_{i,\, al}\, C^{R^*}_{j,\, al}\; 
        f^{(c)}_2(x_{al})\, \bigg], \\
    a_R^{(X)\, ji} & = & \left. a_L^{(X)\, ji}\right|_{L \leftrightarrow R},
\eea
with $x_{al}=m_{\chi^0_a}^2/m_{\tilde \ell_l}^2$ in $f_{1,2}^{(n)}(x)$ and $x_{al}=m_{\chi^-_a}^2/m_{\tilde \nu_l}^2$ in $f_{1,2}^{(c)}(x)$.
The loop functions $f_{1,2}^{(c),(n)}(x)$ are compiled in Appendix \ref{appsec:loop}.
If $i=j=2$, the amplitudes in \Eq{appeq:gmu} are
\bea
    a_L^{(n)\,\mu} & = & -\frac{1}{32\pi^2} \sum_{l,a}\, \frac{1}{m_{\tilde \ell_l}^2} 
        \bigg[\,2\, N^L_{\mu,\, al}\, N^{L^*}_{\mu,\, al}\; f^{(n)}_{1}(x_{al}) \:+\: 
        \frac{m_{\chi^0_a}}{m_\mu}\, N^L_{\mu,\, al}\, N^{R^*}_{\mu,\, al}\; 
        f^{(n)}_2(x_{al})\, \bigg], \label{appeq:aLn_g2} \\
    a_L^{(c)\,\mu} & = & \frac{1}{32\pi^2} \sum_{l,\,a}\, \frac{1}{m_{\tilde \nu_l}^2} 
        \bigg[\,2\,  C^L_{\mu,\, al}\, C^{L^*}_{\mu,\, al}\; f^{(c)}_1(x_{al}) \:+\: 
        \frac{m_{\chi^-_a}}{m_\mu}\, C^L_{\mu,\, al}\, C^{R^*}_{\mu,\, al}\; 
        f^{(c)}_2(x_{al})\, \bigg], \\
    a_R^{(X)\,\mu} & = & \left. a_L^{(X)\, \mu}\right|_{L \leftrightarrow R}.
\eea
First, we compute the chargino amplitude.
Inspecting \Eqs{appeq:CL}{appeq:CR}, one may observe that
\bea
     C_i^L\, C_j^L \propto g_2^2\, m_{\ell_i}\, m_{\ell_j} & \quad\qquad & \frac{m_{\chi_a^-}}{m_{\ell_j}}\, C_i^L\, C_j^R \propto g_2^2\, m_{\chi^-_a} \frac{m_{\ell_i}}{m_{\ell_j}} \\
     C_i^R\, C_j^R \propto g_2^2 & \quad\qquad & \frac{m_{\chi_a^-}}{m_{\ell_j}}\, C_i^R\, C_j^L \propto g_2^2\, m_{\chi^-_a} \frac{\cancel{m_{\ell_j}}}{\cancel{m_{\ell_j}}}
\eea
and anticipate that the leading contributions will come from $C^R_\e\, C^L_\mu$ in $\mu\to\e\gamma$ and from $C^L_\mu\, C^R_\mu$ in $\gmu$.
We expand those terms following the MIA guidance.
As we are not considering complex phases,
\bea
    a_R^{(c)\, \mu\e} & \simeq & \frac{\alpha_2}{4\pi m_w\cos\beta}\; \sum_{l,\,a}\, 
        U_{l\mu}^\nu\, (O_R)_{a1}\, \frac{m_{\chi_a^-}}{m_{\tilde\nu_i}^2}\, 
        f_2^{(c)}(x_{al})\, (O_L)_{a2}\, U_{l\e}^\nu \\
                      & \equiv & \frac{\alpha_2}{4\pi m_w\cos\beta}\; \sum_{l}\, 
        U_{l\mu}^\nu\, F(m_{\chi_a^-}, m_{\tilde \nu_l}^2)\, U_{l\e}^\nu \\
    a^{(c)}_\mu & \simeq & -\frac{\alpha_2\, m_\mu^2}{4\pi m_w\cos\beta}\; 
        \sum_{l,\,a}\,  U_{l\mu}^\nu\, (O_R)_{a1}\, \frac{m_{\chi_a^-}}{m_{\tilde\nu_i}^2}\, 
        f_2^{(c)}(x_{al})\, (O_L)_{a2}\, U_{l\mu}^\nu \\
                            & \equiv & -\frac{\alpha_2\, m_\mu^2}{4\pi m_w\cos\beta}\; \sum_{l}\, U_{l\mu}^\nu\, F(m_{\chi_a^-}, m_{\tilde \nu_l}^2)\, U_{l\mu}^\nu
\eea
Applying the MIA, \Eq{appeq:FET}, for the sneutrino propagator and recalling that in our models $m_{\tilde \nu_\e}\simeq m_{\tilde \nu_\mu}$:
\bea
    a_R^{(c)\, \mu\e} & \simeq & \frac{\alpha_2}{4\pi m_w\cos\beta}\; \left(M_{\tilde 
        \ell_L}^2\right)_{21}\, \left[F(m_{\chi_a^-}, m_{\tilde \nu_l}^2)\right]_{21} \\
                      & \approx & \frac{\alpha_2}{4\pi m_w\cos\beta}\; \frac{\left(M_{\tilde 
        \ell_L}^2\right)_{21}}{m_{\tilde \nu}^4}\; \sum_{a}\, (O_R)_{a1}\; 
        f_3^{(c)}(x_{a\tilde{\nu}})\; (O_L)_{a2}\,  m_{\chi_a^-}, \\
    a^{(c)}_\mu & \simeq & -\frac{\alpha_2\, m_\mu^2}{4\pi m_w\cos\beta}\; \left[F(m_{\chi_a^-}, 
        m_{\tilde \nu_l}^2)\right]_{22} \\
                & \approx & -\frac{\alpha_2}{4\pi m_w\cos\beta}\; 
        \frac{m_\mu^2}{m_{\tilde \nu_\mu}^2}\; \sum_{a}\, (O_R)_{a1}\; f_2^c(x_{a\tilde{\nu}_\mu})\; (O_L)_{a2}\, m_{\chi_a^-},
\eea
with $f_3^{(c)}(x) \equiv f_2^{(c)}(x) + x f_2^{(c)\, '}(x)$ explicitly given in Appendix \ref{appsec:loop}.
We repeat the same procedure for the chargino propagator taking into account the derived equality in \Eq{appeq:FET_fermions}.
Then,
\bea
    a_R^{(c)\, \mu\e} & \simeq & \frac{\alpha_2}{4\pi m_w\cos\beta}\; 
        \frac{\left(M_{\tilde \ell_L}^2\right)_{21}}{m_{\tilde \nu}^4}\; \sum_{a}\, 
        \left(M_C\right)_{1a}\, \left[f_3^{(c)}(x_{a\tilde{\nu}_\mu})\right]_{a2} \nn \\
                     & \approx & \frac{\alpha_2}{4\pi}\; 
        \frac{\left(M_{\tilde \ell_L}^2\right)_{21}}{m_{\tilde \nu}^4}\; \frac{M_2\, 
        \mu}{M_2^2 \,-\, \mu^2}\, F_3^{c}(x_{2\tilde{\nu}},x_{\mu\tilde{\nu})}\, 
        \tan\beta. \label{appeq:mey_MIA} \\
    a^{(c)}_\mu & \simeq & -\frac{\alpha_2}{4\pi m_w\cos\beta}\, 
        \frac{m_\mu^2}{m_{\tilde \nu_\mu}^2}\; \sum_{a}\, \left(M_C\right)_{1a}\, 
        \left[f_2^{(c)}(x_{a\tilde{\nu}_\mu})\right]_{a2} \nn \\
                            & \approx & -\frac{\alpha_2}{4\pi}\,
        \frac{m_\mu^2}{m_{\tilde\nu_\mu}^2}\, \frac{M_2\, \mu}{M_2^2 \,-\, \mu^2}\, 
        F_2^{c}(x_{2\tilde{\nu}_\mu},x_{\mu\tilde{\nu}_\mu)}\, \tan\beta.  \label{appeq:g2_MIA}  
\eea
where in the last equalities we have neglected those terms proportional to $\cos\beta$.
The functions $F_X^{c}(x_1,x_2)$ are defined as $F_X^{c}(x_1,x_2)\equiv f_X^{(c)}(x_1)-f_X^{(c)}(x_2)$.

We can repeat the same procedure for the neutralino amplitude.
From \Eqs{appeq:NL}{appeq:NR} and Eqs.~ \eqref{appeq:aLn_mey} and \eqref{appeq:aLn_g2}, we observe that the dominant term for $\mu\to\e\gamma$ comes from the combination $N^L_\mu N^R_\e$ whereas, in the case of $\gmu$, it is produced by $N^L_\mu N^R_\mu$.
Then, following the same steps than for the chargino loop, one would obtain:
\bea
    a_R^{(n)\, \mu\e} & \simeq & -\frac{\alpha_2}{8\pi}\,
        \frac{\left(M_{\tilde \ell_L}^2\right)_{21}}{m_{\tilde \nu_\mu}^4}\, \frac{M_2\, \mu}{M_2^2 \,-\, \mu^2}\, 
        F_3^{n}(x_{2\tilde{\nu}_\mu},x_{\mu\tilde{\nu}_\mu})\, \tan\beta, \\
    a^{(n)}_\mu & \simeq & -\frac{\alpha_2}{8\pi}\,
        \frac{m_\mu^2}{m_{\tilde \mu}^2}\, \frac{M_2\, \mu}{M_2^2 \,-\, \mu^2}\, 
        F_2^{n}(x_{2\tilde{\nu}_\mu},x_{\mu\tilde{\nu}_\mu})\, \tan\beta,
\eea
with $F_X^n(x_1,x_2)=f_X^{(n)}(x_1)-f_X^{(n)}(x_2)$ and $f_3^{(n)} \equiv f_2^{(n)}(x)+x f_2^{(n)'}(x)$ explicitely given in Appendix \ref{appsec:loop}.
The result is identical to the chargino contribution, except for an additional $1/2$ factor and the loop functions involved.
In a generic model where neutralinos and charginos have similar masses, both amplitudes are equally important.
\begin{figure}[t!]
  \centering
    \begin{minipage}[c]{\textwidth}
        \centering
        \includegraphics[width=0.475\textwidth]{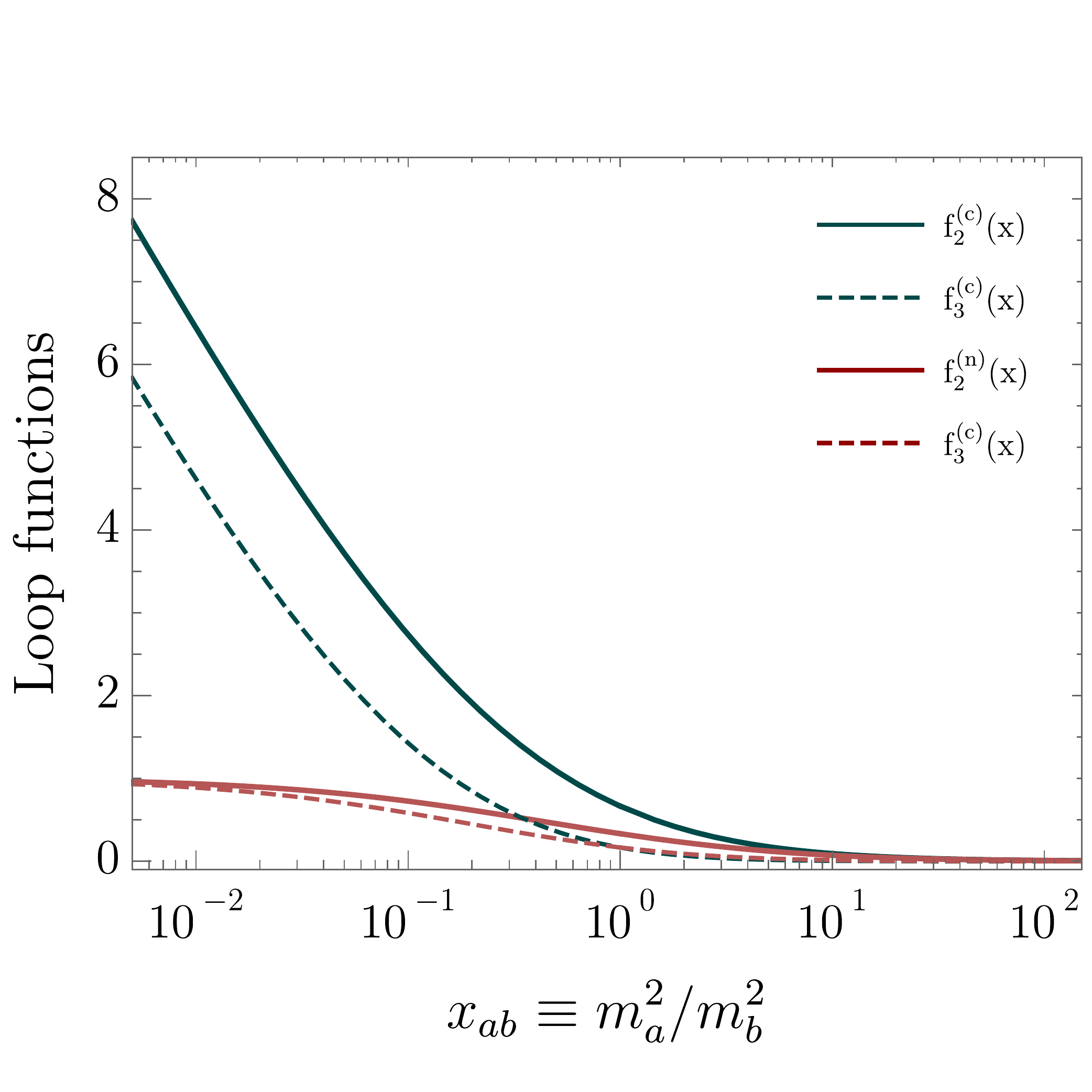}
        \hspace{0.5cm}
        \includegraphics[width=0.475\textwidth]{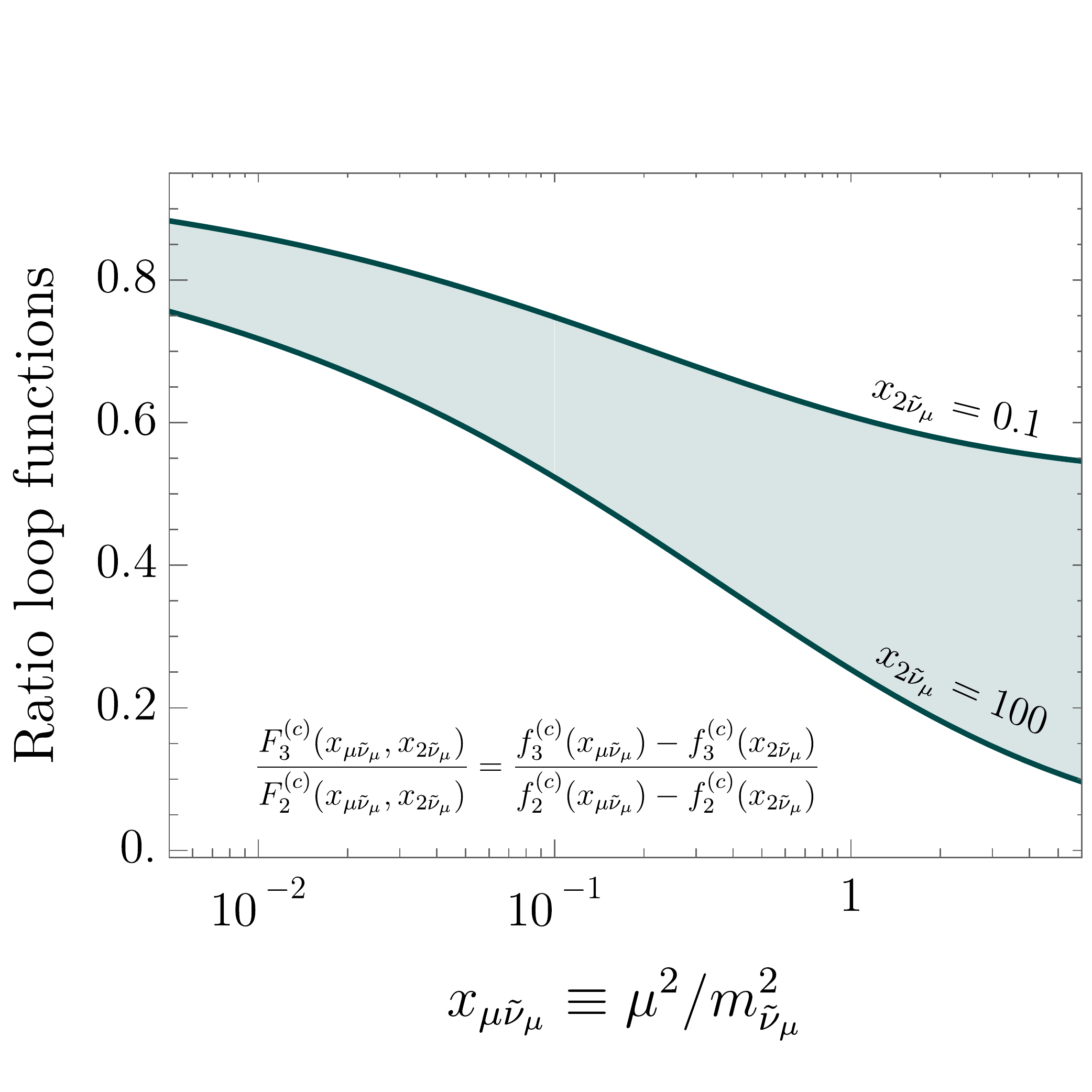}
    \end{minipage}\vspace{-0.cm}\par\vspace{\baselineskip} 
    \captionsetup{width=\textwidth, labelsep=none}
    \caption{\label{appfig:loop} 
        .- Loop functions involved in $a_\mu^{\rm susy}$ (solid line) and ${\rm BR}(\mu\to \e)$ (dashed) corresponding to charginos (green/dark) and neutralinos (red/light).}
\end{figure}

In the models considered here, however, we notice the following hierarchies between masses in the region where $\gmususy$ is within the $3\sigma$ range: $\mu \ll M_2 \lesssim m_{\tilde \mu},\, m_{\tilde \nu_\mu}$.
Evaluating the chargino and neutralino loop functions, \Fig{appfig:loop}, one may see that the former dominates for $x=\mu^2/m_{\tilde \nu_\mu}^2\ll 1$.
Therefore, we expect that the total amplitudes are governed by the chargino terms given in \Eqs{appeq:mey_MIA}{appeq:g2_MIA}.
In \Fig{fig:amu_mey_comp}, we compared the approximate expressions against the exact results obtained with \texttt{SPheno}.
Both cases display a good agreement for the region of interest.
Deviations from the exact result are observed only for small values of $\gmu$ and the region of large ${\rm BR}(\mu\to\e\gamma)$, where other loops may compete and become important.

\section{Majorana mass matrix}
\label{appsec:Majorana}
Here we detail how the Majorana mass matrix can be extracted once the neutrino Dirac matrix is fixed.
In the CKM-like case, the neutrino Yukawa matrix is symmetric and can be expressed as:
\beq \label{appeq:Ynuckm}
    \yckm \;=\; Y_u \;=\; \vckm^T\, Y_u^{\rm diag}\, \vckm.
\eeq
A $M_R$ mass matrix that correctly reproduces the light neutrino masses and mixing can be inferred from Eq.\eqref{eq:seesaw1} as:
\beq \label{appeq:MRckm}
    M_R \;=\; \vu^2~~ \yckm\, (\vpmns\,  m_\nu^{\rm diag^{-1}}\, \vpmns^T)\, \yckm,
\eeq
with $\yckm$ fixed by Eq.\eqref{appeq:Ynuckm}.
In the PMNS-like case, the neutrino Yukawa matrix is given by:
\beq \label{appeq:Ynupmns}
    \ypmns \;=\; Y_u^{\rm diag}\, \vpmns^T.
\eeq
The $M_R$ mass matrix that automatically satisfies all neutrino observables is diagonal and it is simply given by:
\beq \label{appeq:MRpmns}
    M_R \;=\; \vu^2~~ Y_u^{\rm diag}\, m_\nu^{\rm diag^{-1}}\, Y_u^{\rm diag},
\eeq
with eigenvalues
\beq \label{appeq:MRdiagpmns}
    M_R \:=\: {\rm Diag}\left(\cfrac{m_u^2}{m_{\nu_1}},\, \cfrac{m_c^2}{m_{\nu_2}},\, \cfrac{m_t^2}{m_{\nu_3}}\right).
\eeq

\nocite{}
\printbibliography

\end{document}